\documentclass[letterpaper,twocolumn,10pt]{article}
\usepackage{usenix}

\usepackage{tikz}
\usepackage{amsmath}

\usepackage{filecontents}

\usepackage[normalem]{ulem}
\usepackage{acronym}
\usepackage[frozencache,cachedir=.]{minted}
\usepackage{xspace}
\usepackage{pifont}
\usepackage{fontawesome}
\usepackage{multirow}
\usepackage{microtype}
\usepackage{tikz}
\usepackage{enumitem}
\usepackage{tcolorbox}
\newfloat{tcolorboxfloat}{htbp}{lop}
\usepackage{colortbl}
\usepackage{algorithm}
\usepackage{algpseudocode}
\usepackage{pgfplots}
\usetikzlibrary{pgfplots.statistics}
\usepackage{url}
\usepackage{amsmath}
\usepackage{booktabs}
\usepackage{hyperref}

\usepackage{cleveref} %

\begin{document}

\date{}
\newcommand{\red}[1]{{\color{red}#1}}

\newcommand{\datasetTotalTargets}{\red{XXX}} %
\newcommand{\datasetTotalBugs}{\red{XXX}} %
\newcommand{\datasetTotalLeafFunctions}{\red{XXX}} %
\newcommand{\datasetTotalPaths}{\red{XXX}} %
\newcommand{\datasetTotalCallEdges}{166} %

\newcommand{\setupFuzzingTime}{\red{6} hours} %
\newcommand{\setupConstraintRun}{30 minutes} %
\newcommand{\setupStitchingTime}{1 minute} %

\newcommand{\rqoneMatchedTargetFunctionsHigh}{21} %
\newcommand{\rqoneMatchedTargetFunctionsMed}{12} %
\newcommand{\rqoneMatchedTargetFunctionsLow}{6} %
\newcommand{\rqoneMatchedCallEdges}{165} %
\newcommand{\rqoneMatchedTargetFunctionsPercentage}{47.82\%} %
\newcommand{\rqoneMatchedCallEdgesPercentage}{99.39\%} %
\newcommand{\rqoneMatchedCallEdgesWithoutPromotion}{128} %
\newcommand{\rqoneMatchedCallEdgesWithoutPromotionPercentage}{77.10\%} %
\newcommand{\rqoneSampleMatchedEdgesChangePercentage}{\red{XXX}\%} %
\newcommand{\rqoneHarnessesGenerated}{\red{XXX}} %
\newcommand{\rqoneArgumentsGenerated}{\red{XXX}} %
\newcommand{\rqoneGlobalsGenerated}{\red{XXX}} %
\newcommand{\rqoneArraysGenerated}{\red{XXX}} %
\newcommand{\rqoneArrayRelationsGenerated}{\red{XXX}} %
\newcommand{\rqoneAverageArgumentsPerHarness}{\red{XXX}} %
\newcommand{\rqoneAverageGlobalsPerHarness}{\red{XXX}} %
\newcommand{\rqoneRandomSampledHarnesses}{\red{XXX}} %
\newcommand{\rqoneTotalCrashes}{338} %
\newcommand{\rqoneTotalConstraintsGenerated}{284} %
\newcommand{\rqoneTotalConstraintsGeneratedPercentage}{84\%} %
\newcommand{\rqoneRandomSampledFailedConstraint}{\red{XXX}} %
\newcommand{\rqoneRandomSampledSuccessfulConstraint}{\red{XXX}} %
\newcommand{\rqoneLeafLevelCrashes}{\red{XXX}} %
\newcommand{\rqoneLeafLevelConstraints}{\red{XXX}} %
\newcommand{\rqoneLeafLevelCrashesPercentage}{\red{XXX}\%} %
\newcommand{\rqoneCallEdgeCrashes}{204} %
\newcommand{\rqoneCallEdgeConstraints}{193} %
\newcommand{\rqoneCallEdgeCrashesPercentage}{94.61\%} %
\newcommand{\rqoneOutOfBoundPointersPercentage}{\red{XXX}\%} %
\newcommand{\rqoneDidnotReachPercentage}{\red{XXX}\%} %
\newcommand{\rqoneDidnotFinishPercentage}{\red{XXX}\%} %
\newcommand{\rqoneTotalCountRemovedVariables}{\red{XXX}} %
\newcommand{\rqoneNullPointerPercentage}{\red{XX}\%} %
\newcommand{\rqoneTotalSizeReductionsPercentage}{43\%} %
\newcommand{\rqoneRandomSampledFailedStitching}{\red{XXX}} %
\newcommand{\rqoneArrayRelationsIdentified}{\red{XXX}}
\newcommand{\rqDidnotFinishPercentage}{\red{XXX}}

\newcommand{\rqtwoTPCount}{28} %
\newcommand{\rqtwoFPCount}{6} %
\newcommand{\rqtwoPrecision}{100\%} %
\newcommand{\rqtwoFNCount}{20} %
\newcommand{\afgentpCount}{33} %
\newcommand{\afgenPrecision}{94\%} %
\newcommand{\afgenRecall}{\red{XXX}} %
\newcommand{\afgenfpRate}{\red{YY}\%} %
\newcommand{\rqtwoWeFoundAFGenDidntBugs}{2} %
\newcommand{\rqtwoCoveragePercentage}{\red{13}\%} %
\newcommand{\rqtwoNoCrashesForTargetFunctions}{13} %
\newcommand{\rqtwoNoCrashesForTargetFunctionsPercentage}{28\%} %
\newcommand{\rqtwoTotalUniqueLeafCrashes}{37} %
\newcommand{\rqtwoTotalRequiredCrashes}{48} %
\newcommand{\rqtwoNullPtrCrashesPercentage}{\red{XX}\%} %
\newcommand{\rqtwoAverageTimeToFindCrashes}{\red{XXX}} %
\newcommand{\rqtwoAverageTimeToGenerateConstraints}{\red{XXX}}
\newcommand{\rqtwoDidNotReachPercentage}{\red{XX}\%} %
\newcommand{\rqtwoReachedDidNotTriggerPercentage}{\red{XX}\%} %
\newcommand{\rqtwoCallEdgesTotal}{\red{214}} %
\newcommand{\rqtwoCallEdgesCountingCallsitesTotal}{\red{XXX}} %
\newcommand{\rqtwoCallEdgesReached}{190} %
\newcommand{\rqtwoCallEdgesReachedPercentage}{89\%} %
\newcommand{\rqtwoCallEdgesConstraints}{178} %
\newcommand{\rqtwoCallEdgesConstraintsPercentage}{94\%} %
\newcommand{\rqtwoRandomSampledNotReached}{\red{XXX}} %

\newcommand{\rqtwoMaxCoveragePercentage}{\red{XXX}}
\newcommand{\rqtwoMinCoveragePercentage}{\red{XXX}}
\newcommand{\rqtwoTotalUniqueLeafCrashesPercent}{72\%}
\newcommand{\rqtwoCallEdgesMatched}{\red{XXX}}
\newcommand{\rqtwoCallEdgesMatchedPercentage}{\red{XXX}}

\newcommand{\rqfourbugschallenging}{\red{XXX}}
\newcommand{\rqfourbugsossfuzz}{\red{XXX}}

\newcommand{\ourtool}{{\sf Griller}}
\newcommand{\sht}{{\emph Staged Hybrid Testing}}
\newcommand{\tooltitle}{{\sf Griller}}
\newcommand{\klee}{{\sf KLEE}}

\newcommand{\sid}[1]{\textcolor{red}{\textbf{Sid : #1}}}

\newcommand{\machiry}[1]{\textcolor{green}{\textbf{Machiry : #1}}}

\newcommand{\arjun}[1]{\textcolor{blue}{\textbf{AA : #1}}}
\newcommand{\code}[1]{%
  \mintinline[fontsize=\small{},mathescape, escapeinside=||]{c}{#1}%
}  

\newcommand\encircle[1]{\microtypesetup{deactivate}\tikz[baseline=(X.base)] \node (X) [draw, shape=circle, inner sep=0em,text width=1em, text centered] {\scriptsize #1};\microtypesetup{reactivate}}

\newcommand{\eg}{\textit{e.g.,}\xspace}
\newcommand{\ie}{\textit{i.e.,}\xspace}
\newcommand{\etal}{\textit{et al.}\xspace}

\newcommand{\tbl}[1]{Table~\ref{#1}}
\newcommand{\sect}[1]{\S~\ref{#1}}
\newcommand{\apdx}[1]{\S~\ref{#1}}
\newcommand{\fig}[1]{Figure~\ref{#1}}
\newcommand{\lst}[1]{Listing~\ref{#1}}
\newcommand{\algo}[1]{Algorithm~\ref{#1}}
\newcommand{\tcrash}{\textcolor{red}{\faBug}}
\newcommand{\fcrash}{\textcolor{green}{\faBug}}
\newcommand{\grilend}{\textcolor{green}{\faTimesCircle}}
\newcommand{\traceback}{\textcolor{red}{\faArrowCircleUp}}

\acrodef{FLT}{Function-Level Testing}
\acrodef{PUT}{Program Under Test}
\newcommand{\bout}{{\sc Bout}\xspace}
\newcommand{\rbout}{{\sf Reactive Bottom-Up Testing}\xspace}

\def\fullsupp{\ding{51}}
\def\partsupp{\ding{109}}
\def\nosupp{\ding{55}}

\definecolor{americanrose}{rgb}{1.0, 0.01, 0.24}
\definecolor{antiquefuchsia}{rgb}{0.57, 0.36, 0.51}
\definecolor{asparagus}{rgb}{0.53, 0.66, 0.42}
\definecolor{celadon}{rgb}{0.67, 0.88, 0.69}
\definecolor{jade}{rgb}{0.0, 0.66, 0.42}
\definecolor{lightseagreen}{rgb}{0.13, 0.7, 0.67}
\definecolor{persiangreen}{rgb}{0.0, 0.65, 0.58}
\definecolor{puce}{rgb}{1.0, 0.53, 0.6}
\definecolor{sacramentostategreen}{rgb}{0.0, 0.34, 0.25}
\definecolor{chartreuse}{rgb}{0.5, 1.0, 0.0}
\definecolor{uclagold}{rgb}{1.0, 0.7, 0.0}
\definecolor{uscgold}{rgb}{1.0, 0.8, 0.0}
\definecolor{tigerseye}{rgb}{0.88, 0.55, 0.24}
\definecolor{britishracinggreen}{rgb}{0.0, 0.26, 0.15}
\definecolor{khaki}{rgb}{0.76, 0.69, 0.57}
\definecolor{lavender}{rgb}{0.71, 0.49, 0.86}
\definecolor{maroon}{rgb}{0.69, 0.19, 0.38}
\definecolor{cadetblue}{rgb}{0.37, 0.62, 0.63}
\definecolor{caribbeangreen}{rgb}{0.0, 0.8, 0.6}
\definecolor{aliceblue}{rgb}{0.94, 0.97, 1.0}
\acrodef{CVSS}{Common Vulnerability Scoring System}
\acrodef{IPC}{Inter-process communication}

\newcommand{\dr}{\ensuremath{P_{d}}}
\newcommand{\inc}{\ensuremath{C^{r}}}
\newcommand{\scr}{\ensuremath{c^{r}}}
\newcommand{\pcr}{\ensuremath{\varphi}}
\newcommand{\crb}{\ensuremath{C^{r}_{B}}}

\newcommand{\numnewbugs}{6\xspace}

\definecolor{bugtrigerringstates}{rgb}{0.9, 0.17, 0.31}
\definecolor{invalidstates}{rgb}{1.0, 1.0, 0.2}
\definecolor{validstates}{rgb}{0.8, 0.89, 1.0}
\definecolor{infeasiblestates}{rgb}{0.2, 0.6, 1.0}
\definecolor{potentiallybuggystates}{rgb}{1.0, 0.7, 0.4}

\acrodef{TDT}{Top Down Testing}

%\title{\Large \bf Formatting Submissions for USENIX Security 2026:\\
%  An (Incomplete) Example}
\title{\Large Reactive Bottom-Up Testing}
\author{{\rm Siddharth Muralee, Sourag Cherupattamoolayil, James C. Davis,  Antonio Bianchi, Aravind Machiry}\\ 
Purdue University, \{smuralee, scheupa, davisjam, antoniob, amachiry\}@purdue.edu \\
}

\maketitle

\begin{abstract}

Modern computing systems remain rife with software vulnerabilities.
Engineers apply many means to detect them, of which dynamic testing is one of the most common and effective.
However, most dynamic testing techniques follow a top-down paradigm, and struggle to reach and exercise functions deep within the call graph.
While recent works have proposed Bottom-Up approaches to address these limitations,
they face challenges with false positives and generating valid inputs that adhere to the context of the entire program.

In this work, we introduce a new paradigm that we call Reactive Bottom-Up Testing.
Our insight is that function-level testing is necessary but not sufficient for the validation of vulnerabilities in functions.
What we need is a systematic approach that not only tests functions in isolation but also validates their behavior within the broader program context, ensuring that detected vulnerabilities are both reachable and triggerable.
We develop a three-stage bottom-up testing scheme: (1) identify likely-vulnerable functions and generate type- and context-aware harnesses; (2) fuzz to find crashes and extract input constraints via symbolic execution; (3) verify crashes by combining constraints to remove false positives.
We implemented an automated prototype, which we call \ourtool{}.
We evaluated \ourtool{} in a controlled setting using a benchmark of 48 known vulnerabilities across 5 open-source projects, where we successfully detected 28 known vulnerabilities.
Additionally, we evaluated \ourtool{} on several real-world applications such as Pacman, and it discovered 6 previously unknown vulnerabilities.
Our findings suggest that Reactive Bottom-Up Testing can significantly enhance the detection of vulnerabilities in complex systems, paving the way for more robust security practices.

\end{abstract}

\section{Introduction}
\label{sec:intro}

Software testing~\cite{myers2011art, ammann2016introduction} is an important capability to find security vulnerabilities.
There are many automated software testing approaches with varying capabilities and customizations for different types of software, \eg{}
  network programs~\cite{qin2023nsfuzz, schumilo2022nyx,gorbunov2010autofuzz, natella2022stateafl}, 
  web APIs~\cite{godefroid2020intelligent,hatfield2022deriving, atlidakis2019restler},
  and
  autonomous vehicles~\cite{zhong2022neural,moukahal2021vulnerability,kim2022drivefuzz}.
Most existing techniques follow a top-down paradigm, where the \ac{PUT} is exercised from its entry point (\eg{} \code{main}) with the goal of execution reaching an invalid state (\ie{} a bug) and/or improving testing coverage.
Top-down approaches struggle to reach deeper program points~\cite{bohme2020fuzzing, gao2023beyond} and generate inputs that can provide full coverage~\cite{wang2019neufuzz, she2019neuzz}.
Directed fuzzing~\cite{bohme2017directed, 10.1145/3238147.3238176} tries to tackle this by focusing on reaching specific functions, but they still suffer from reaching deeper functions.

To address these inherent limitations of top-down exploration, the Bottom-up Testing paradigm (\bout{})\cite{ISO24765:2010} proposes analyzing the \ac{PUT} directly at arbitrary program points.
The primary challenge to \bout{} is \emph{feasible states generation}.
Specifically, to avoid false-positives (\eg{} infeasible argument values), a \bout{} approach should be aware of the feasible program states that can reach the target function and should have the ability to generate those.
Existing techniques take a proactive approach to tackle the feasible states problem.
However, they require considerable manual effort\cite{bohme2020fuzzing, mallissery2023demystify}.
\emph{In addition to manual effort, the proactive approach is wasteful if the target function is bug-free.}

As a solution, in this paper we propose \emph{a reactive approach}.
With this approach, instead of generating only feasible states, we check for the validity of bug-triggering states.
We call this approach \rbout{}. 
We exercise the target function in an unconstrained manner and identify states resulting in bugs.
Then, we check the feasibility of only these bug-triggering states in \ac{PUT}.
We realized our approach by implementing \ourtool{}, a framework enabling automated and efficient bottom-up testing.
We use a combination of symbolic execution and constraint stitching to realize our reactive approach.
Specifically, we symbolically trace each crashing input and collect the crash constraint,~\ie{} a symbolic constraint that, if satisfied by the arguments, triggers the crash.
Given a crash constraint, we backtrack along the original program call graph through fuzzing and symbolic tracing.
We check whether the crash is feasible by stitching the crashing constraint with the symbolic value of arguments at each call site.
We use a \emph{staged approach for backtracking} and save the symbolic state along each call edge.
This enables us to \emph{reuse common backtracking information across multiple functions and avoid unnecessary exploration}.

\ourtool{} works in three distinct phases: 
First, \emph{target function identification and harness generation}, wherein points-of-entry functions with high vulnerability potential are algorithmically identified and isolated for independent testing.
We identify argument type through static analyses and type inference, and generate the harness through instrumentation.
Second, \emph{constraint extraction} through symbolic execution, whereby crashes identified during function-level fuzzing are analyzed to extract the constraints characterizing the vulnerability and the execution path leading to it. 
Finally, \emph{infeasible crash filtering} through backtracking and constraint stitching, which determines whether the extracted constraints are satisfiable within the context of the entire application.

We evaluate \ourtool{} on a dataset of 48 CVEs from 5 open source projects and show that it can successfully find vulnerabilities comparable with other techniques.
\ourtool{} successfully identified vulnerabilities that existing state-of-the-art tools failed to detect, demonstrating complementary detection capabilities.
We also found \numnewbugs{} previously unknown vulnerabilities in large real-world programs, including Pacman (package manager for Arch Linux).

\noindent In summary, our contributions are the following:

\begin{itemize}
\item We introduce the Reactive Bottom-Up Testing paradigm, a new approach to testing that begins by identifying sensitive functions, testing them in isolation and filtering infeasible states in a reactive manner.
\item We present the first implementation of Reactive Bottom-Up Testing paradigm through a prototype called Griller. 
We use a blend of fuzzing and symbolic execution to achieve it.
\item We show the potential of the Reactive Bottom-Up Testing paradigm through evaluations on an existing CVE benchmark.
We also ran \ourtool{} on real-world applications, and found \emph{\numnewbugs{} previously unknown (zero-day) security vulnerabilities} among which \emph{4 have been acknowledged and patched}.
\end{itemize}

\section{Background}

\subsection{Top-Down Testing}
\label{sec:topdowntestingoverview}

\begin{figure}[t]
\includegraphics[scale=0.65]{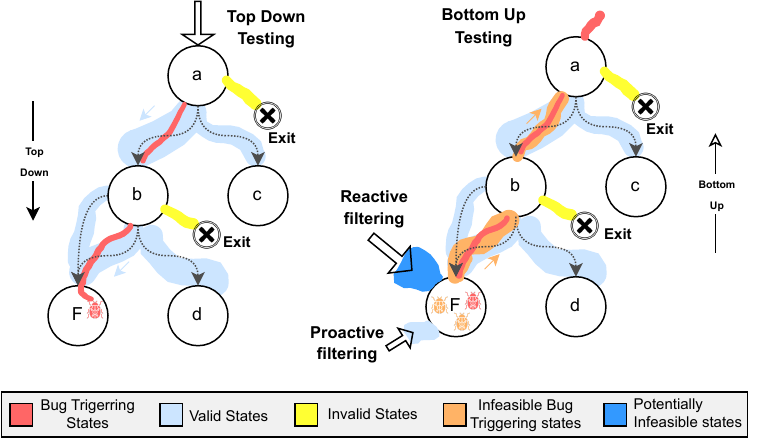}
\caption{
Difference between \ac{TDT} and \bout{} testing approaches. 
}
\label{fig:bottomuptesting}
\end{figure}

Most common testing techniques are traditionally performed in a top-down manner along the \acf{PUT}'s call-graph, as shown in \fig{fig:bottomuptesting}.
Specifically, given a \ac{PUT}, the \ac{TDT} technique executes the \ac{PUT} and provides input through required channels, \eg{} standard input, files, sockets, etc.
The \ac{PUT} sets up the \colorbox{validstates}{states} along the execution flow needed for subsequent functions.
The \ac{PUT} might constrain the state space at arbitrary program points (\eg{} through \code{if} statements), \ie{} consider certain states as invalid and terminate execution (as \colorbox{invalidstates}{illustrated} by \fig{fig:bottomuptesting}).
Consequently, the state space reachable at deeper (\eg{} along the \ac{PUT}'s call graph) program points gets progressively reduced.
Certain states might result in unexpected program behaviors, \eg{} \code{SEGFAULT}, Integer overflow, etc.; some might be security vulnerabilities.
\ac{TDT} approaches (\eg{} Fuzzing \cite{manes2019art}) try to generate inputs that make \ac{PUT} generate states that can maximize certain goal(s), \eg{} to reach deeper valid program points (improve coverage), trigger certain unexpected behaviors (\eg{} buffer-overflows, race-conditions).

One of the \emph{disadvantages of \ac{TDT} is unnecessary state exploration}. 
As shown in \fig{fig:bottomuptesting}, despite only limited interesting states (\eg{} \colorbox{bugtrigerringstates}{bug-triggering states}), \ac{TDT} approaches generate unnecessary state space.
This is a fundamental limitation; as \ac{TDT} approaches exercise \ac{PUT} from the top, \ie{} program entry point, it is hard to know which state would trigger interesting states at later execution points.
Some techniques \cite{wang2020not} try to tackle this by prioritizing inputs that are more likely to generate interesting states. However, these are specialized for specific classes of programs and are hard to generalize.
Another disadvantage is unnecessary function exploration.
As explained in \sect{sec:intro}, not all functions are relevant from a security perspective.
For instance, in a web server application, functions that parse user input have a significantly higher security risk.
Though some directed testing techniques~\cite{bohme2017directed, 10.1145/3238147.3238176} try to tackle this by focusing on inputs reaching a specific function, they still suffer from unnecessary state exploration~\cite{wang2024progress}.

\subsection{Bottom-Up Testing}
\label{sec:bottomuptestingoverview}

Another proposed testing approach is Bottom-Up Testing (\bout{})~\cite{ISO24765:2010}.
In this approach, function of interest is known (\eg{} $F$ in \fig{fig:bottomuptesting}), and is directly tested by invoking it with appropriate arguments.
However, function could be excercised with \colorbox{infeasiblestates}{infeasible states}, \ie{} those that are not feasible in the \ac{PUT}.
If these states do not lead to incorrect behavior(bugs), then these infeasible states do not matter, and in fact, get higher confidence in the security of the function.
But, \emph{if any of these states trigger bugs, they could be false positives unless it can show that the corresponding states are feasible in \ac{PUT}.}
Specifically, \bout{} approaches need to assess whether bug triggering states are feasible (\ie{} reachable along the control flow) from the program entry point.
\emph{This avoids the problems of unnecessary state exploration and enables us to focus on specific functions.}

\subsubsection{Proactive Bottom-Up Testing}
Although \bout{} has been proposed for some time, its practical adoption has been limited, primarily due to the difficulty of identifying and managing infeasible states.
One of the ways to avoid infeasible states is to filter them \emph{proactively} while exercising the function.
Unit testing harnesses~\cite{khorikov2020unit} are a classic example of this approach, where the developer creates appropriate and feasible states (\eg{} through mock objects \cite{de2023mock}) for a function before invoking it directly.
However, there is considerable manual effort \cite{nourry2023human} involved in creating effective harnesses~\cite{spadini2017mock, spadini2019mock}.
Although some techniques, such as AFGen~\cite{liu2024afgen, sherman2025no}, try to automate it, they are domain-specific and require examples.
These approaches more closely resemble constrained top-down testing techniques than pure bottom-up testing.
For instance, AFGen generates test harnesses that must be valid within the application context.
These harnesses contain code snippets (including control statements) taken from the original program that are used to create the necessary context for testing the target function.
These are similar to directed fuzzing approaches like Beacon~\cite{9833751}, where constraints along the path are promoted to direct the fuzzing process towards a program point of interest.
If the generated harnesses are inherently complex, then they suffer from the same reachability issues as \ac{TDT} approaches.

Instead, in this paper, we follow a reactive approach and check for the feasibility of only those states we know are bug-triggering.
The idea is to test the function in an unconstrained manner and identify \colorbox{potentiallybuggystates}{bug-triggering states}.
Then, the feasibility of these states is checked along the program call graph.

\subsection{Terms and Notations}
\label{sec:terms}

\hspace*{0.9em} \textbf{Target Function ($f$)}: A specific function within the \ac{PUT} selected as the entry point for testing.
For instance, \code{add_elem} in \lst{lst:runexample}(a).
%\\

\vspace{0.3em}
\indent \textbf{Call Edge ($ei$)}: An ordered pair $ei=(pf, cf)$, where $pf$ denotes the caller function and $cf$ represents the callee function such that there exists at least one call of $cf$ in $pf$, \eg{} \code{(process_req, add_elem)} in \lst{lst:runexample}.
Multiple invocation sites of $cf$ within $pf$ are abstracted as a single call edge in our representation. 
The set of all call edges forms the program's call graph.
% \\%\\

\vspace{0.3em}
\indent \textbf{Driver Program ($\dr$)}: A modified version of \ac{PUT} for a function $f$.
Specifically, $\dr^{f}$ is a modified version of \ac{PUT} that considers $f$ as the entry point and invokes it with arguments of appropriate type created using the standard input.
\lst{lst:trampoline} shows the driver program for \code{add_elem} in \lst{lst:runexample}(a).
% \\%\\

\vspace{0.3em}
\indent \textbf{Crashes ($\inc_f$)}:
A set of crash pairs $\inc_{f}=\{(I_1, c_1) ..., (I_n, c_n)\}$, where each pair $(I_i, c_i)$ indicates that the driver program $\dr^{f}$ will crash with crash signature($c_i$) when executed with input $I_i$. 
The crash signature contains information about the crash location and type (\eg{} buffer overflow).
% \\%\\
% \\%\\

\vspace{0.3em}
\indent \textbf{Root-cause Constraint ($rc_i$)}:
The specific condition at a program point that directly triggers crash $c_i$ when violated (e.g., pointer should be null pointer, or index should be out of bounds for a buffer access).
% \\%\\

\vspace{0.3em}
\indent \textbf{Crashing Constraints ($\scr_i$)}:
A symbolic constraint that, when satisfied by an input, results in crash $c_i$.
It is composed as $\pcr_i \land rc_i$, where $\pcr_i$ represents the path constraints of the crash and $rc_i$ is triggers the actual violation.
The first column of \tbl{tbl:backtrack} shows crashing constraints for crashes in \lst{lst:runexample}(a).
% \\%\\

\vspace{0.3em}
\indent \textbf{Edge Crashes ($\inc_{ei}$)}:
Subset of Crashes corresponding to a Call Edge($ei$).
Specifically, edge crashes $\inc_{ei}$ are a set of inputs, where each input $I_i$ executes a path to the callsite of $cf$ when the driver program $\dr^{pf}$ is executed.
% \\%\\

\vspace{0.3em}
\indent \textbf{Edge Constraints ($\pcr_j$, $a^{cf}_j$)}:
At each callsite $j \in callsites(pf, cf)$, we extract two critical elements: the path constraint $\pcr_j$ representing the necessary conditions to reach the callsite, and the symbolic values of arguments $a^{cf}_j$ passed to $cf$. 
Together, these form an edge constraint pair that characterizes the conditions for executing the call edge.
Columns 2 and 3 in \tbl{tbl:backtrack} show the edge constraints for the call edge \code{(process_req, add_elem)} in \lst{lst:runexample}(a).
% \\%\\

\vspace{0.3em}
\indent \textbf{Constraint Stitching}:
\label{sec:constraint_stitching}
Constraint Stitching for a target function $f$ and a call edge $(pf, cf)$ is possible if $cf$ = $f$.
Specifically, we create a new crashing constraint: $\pcr_{j} \land (\scr_{i} \odot a^{f}_{j})$, where $\scr_{i} \odot a^{f}_{j}$ is the crashing constraint obtained by replacing the symbolic value of parameters of $f$ with symbolic values of the arguments $a^{f}_{j}$ that are passed at the call-site $j$ and $\pcr_{j}$ represents the path constraint necessary to reach the callsite. 
We refer to this as \emph{Constraint Stitching} or just \emph{Stitching}.

\section{Reactive Bottom-Up Testing}
\label{sec:overview}

\begin{listing*}[t!]
  \begin{tabular}{c|c|c}
    \begin{minipage}{.33\textwidth}
\begin{minted}{c}
struct pool {
 ...
 unsigned size;
 bool avail;
 char *buf;
 char log[1000];
 ...
};

int add_elem(void *data, 
             const char *buf, 
             unsigned size) {
 struct pool *p = (struct pool *)data;
 // c1: NULL-ptr dereference of
 // pointer p
 |\encircle{$c_{1}$}\fcrash|if (p->avail) {
   p->buf = malloc(
             p->size*sizeof(char));
   if (p->buf) {
     p->avail = 0;
     // c2: Buffer overwrite of p->buf 
     // c3: NULL-ptr dereference of 
     //     pointer buf 
     // c4: Buffer overread of buf
     memcpy(|\encircle{$c_{2}$}\tcrash|p->buf, 
            |\encircle{$c_{3}$}\fcrash\encircle{$c_{4}$}\fcrash|buf, size);
   }
   return 0;
 }
 return -1;
}
\end{minted}
    \end{minipage} &
    \begin{minipage}{.27\textwidth}
\begin{minted}{c}
struct pool GP[POOL_SIZE];
enum rtype { ..,NEW, UPDATE,..};
int process_req(enum rtype r, 
                uint32_t idx,
		char *buf, 
		uint32_t sz) {

 if (idx < POOL_SIZE) {
   GP[idx].size = size;
   switch(r) {
     case UPDATE:
       |   \sout{\encircle{$c_{1}$}} \sout{\encircle{$c_{2}$}} \encircle{$c_{3}$} \encircle{$c_{4}$}|
       |$\text{\faArrowRight}_{1}$|add_elem(&GP[idx],
                    buf,
                    sz);
       break;
     ...  
     case NEW:
       |   \sout{\encircle{$c_{1}$}} \encircle{$c_{2}$} \sout{\encircle{$c_{3}$}} \sout{\encircle{$c_{4}$}}|
       |$\text{\faArrowRight}_{1}$|add_elem(&GP[idx],
                    "HDR",
                    3);
       break;
     case DELETE:
       ...
       break; 
   }
 }
 ...
}
\end{minted}
    \end{minipage} &
    \begin{minipage}{.27\textwidth}
\begin{minted}{c}
int handle_req(enum rtype r,
               uint32_t gidx, 
               int sockfd) {
 
 char rbuf[MAX_BUF_SIZE];
 int sz = MAX_BUF_SIZE;
 if (r == NEW) {
  // read from socket
  // Note: sz can be less than 3
  sz = read(sockfd,
            &rbuf,
            sizeof(rbuf));
  if (sz <= 0)
   return -1;
   |  \encircle{$c_{2}$} \sout{\encircle{$c_{3}$}} \sout{\encircle{$c_{4}$}}|
 |$\text{\faArrowRight}_{2}$|process_req(r, gidx, &rbuf, sz);
 }
 ...
}
\end{minted}
    \end{minipage}\\
(a) Target function $f$. & (b) Caller of~\code{add_elem} ($1^{st}$ level).  & (c) Caller of~\code{process_req} ($2^{nd}$ level). \\ 
  \end{tabular}
\caption{Running example demonstrating~\ourtool{} approach. Where~\tcrash{} and~\fcrash{} indicate true (bug-revealing) and false (infeasible) crashes respectively. $\text{\faArrowRight}_{x}$ indicates call-sites at level $x$ that are back traced and corresponding filtered out (\sout{striked}) crashes.}
\label{lst:runexample}
\end{listing*}

In this section, we demonstrate the Reactive \bout{} concept through a concrete example and identify the key technical challenges that must be addressed in any practical implementation. 
We illustrate how our specific instantiation of \bout{} makes design choices for each component of the general framework using a running example.

Consider ~\lst{lst:runexample}, which shows a code snippet with potential vulnerabilities. 
We start fuzzing a \ac{PUT} from functions that are potentially risky, say \code{add_elem} in \lst{lst:runexample} (a), by generating arguments and any used global variables of the appropriate type from standard input.
This might result in crashes ($c_{1}$ - $c_{4}$ in \lst{lst:runexample} (a)), and we capture the corresponding crash-causing states as symbolic constraints (first column in \tbl{tbl:backtrack}).
We check for the feasibility of these constraints from the caller along the call graph by stitching the symbolic value of arguments and path constraints (columns 2 and 3 of \tbl{tbl:backtrack}).
We drop the constraints that are infeasible (\textcolor{red}{\faRemove} in \tbl{tbl:backtrack}).
For the remaining constraints (\textcolor{green}{\faCheck} in \tbl{tbl:backtrack}), we continue this process backward along the call graph (\tbl{tbl:backtrack2}) until the program entry point, \ie{} \code{main}.
We report the crashes corresponding to the remaining constraints as true bugs (\textcolor{green}{\faCheck} in \tbl{tbl:backtrack2}).
In the following sections, we provide an overview of each of our steps and the corresponding challenges.

\subsection{Target Function Identification and Harness Generation}
\label{overview:phaseone}

Given a program $P$, the first component of any \bout{} system must select target functions for analysis.
While various strategies are possible (such as manual selection or random sampling).
Our approach employs heuristics (detailed in \sect{sec:targetidentifier}) to select \code{add_elem} in \lst{lst:runexample} (a) as the target function ($f$).
The driver program ($\dr^{f}$) for $f$, is shown in \lst{lst:trampoline}.
To accurately generate the input space of $f$, we need create data generators for each required type.

However, the construction of a comprehensive yet efficient harness presents several technical challenges. 
First, achieving type and context awareness requires sophisticated program analysis to determine both explicit and implicit type relationships—for instance, in \code{add_elem}, static analysis reveals that the parameter \code{buf} must reference an array of minimum \code{size} elements. 
Second, completeness requires identifying all relevant variables and their types, including globals and those used by callee functions.
For instance, the fact that \code{data} is actually of type \code{struct pool*} despite being declared as \code{void*}. 
Third, maintaining minimality requires excluding irrelevant components from generation, such as the \code{log} field within \code{struct pool}, since increase in the input space will slow down the fuzzing and symbolic execution techniques.
As detailed in \sect{sec:funcfuzz}, our approach addresses these challenges through targeted program analysis techniques.
\begin{tcolorbox}[title=\small{\textbf{Challenge 1: Function-Level Testing Harness}}, colback=gray!30!white,
colframe=black,
top=1pt,
bottom=1pt
]
\small{
Function-level testing requires a harness that satisfies three critical properties:
\emph{(1) Type and Context awareness},
\emph{(2) Completeness}, and
\emph{(3) Minimality}.}
\end{tcolorbox}

\begin{listing}
\begin{minted}{c}
int main(int argc, char **argv) {
  griller_driver(); return 0; 
  // terminate immediately 
}
void *gen_ptr_for_pool() { // for struct pool
 if (gen_null()) return NULL; 
 struct pool *p = malloc(sizeof(pool)); 
 // initialize fields of p with data from stdin
 p->size = gen_unsigned(); ...
}
char *gen_char_ptr_with_size(unsigned sz) {
 if (gen_null()) return NULL;
 char *p = malloc(sz*sizeof(char));
 // initialize each byte of p with data from stdin
}
unsigned gen_unsigned() {
 //read from stdin
 read(stdin, &p, sizeof(p)); return p;
}
bool gen_null() { 
 read(stdin, &byte, sizeof(byte)); return byte > 25; 
}
void griller_driver() {
 void *p1 = gen_ptr_for_pool();  // argument 1
 unsigned p3 = get_unsigned();   // argument 3
 char *p2 = gen_char_ptr_with_size(p3); // argument 2
 add_elem(p1, p2, p3);  // call the target function
}
\end{minted}
  \caption{Driver Program ($\dr$):  Instrumentation done by\\ ~\ourtool{} to facilitate fuzzing of the function~\code{add_elem} directly.}
\label{lst:trampoline}
\end{listing}

\subsection{Constraint Extraction}
\label{overview:phasetwo}

After generating the harnesses, we employ fuzzing to discover potential crashes in the driver program ($\dr^{f}$). 
These crashes, denoted as \encircle{$c_{x}$} in \lst{lst:runexample}(a), represent potential vulnerabilities.
To determine the precise conditions triggering each crash, we perform symbolic execution of $\dr^{f}$ while \emph{pre-constraining} the symbolic input bytes to match the concrete crashing input $I_{i}$.
For instance, in~\lst{lst:runexample} (a) for~\encircle{$c_{1}$}, a~\code{NULL}-ptr dereference, we check if the pointer involved at the crash site~\ie{}~\code{p} is~\code{NULL} and dump the corresponding constraint in terms of symbolic values of the parameters.
Similarly, for~\encircle{$c_{2}$}, we check that size of the memory object  pointed by~\code{p->buf}~\ie{} that ~\code{p->size} is $\leq$~\code{size}.
The first column in \tbl{tbl:backtrack} shows these crashing constraints. 
For example, crash \encircle{$c_2$}'s constraint ((S1 != NULL)$\land$($S2 < S4$)) combines the path constraint (S1 != NULL) with the root-cause constraint ($S2 < S4$).

To generate \emph{Crashing Constraints}, we require each input to follow the same execution path during both fuzzing and symbolic execution phases.
However, these environments differ fundamentally; their input formats are different and their execution environments are also different: fuzzers execute native compiled binaries directly, while symbolic execution engines typically interpret code at higher levels of abstraction (\eg{} LLVM IR) with custom memory models. 
This can lead to inconsistencies in behavior, especially when dealing with undefined behavior in C.
For example, the binary might not crash due to an off-by-one or an overshift error, but the symbolic execution engine will terminate if it encounters these errors.
Additionally, symbolic execution engines impose limitations on supported operations — many cannot handle symbolic array sizes (\code{buf} should be a symbolic array of length \code{size}; where \code{size} is symbolic) or accurately track memory object relationships (such as \code{obj_size(p->buf) == p->size}). 

\begin{tcolorbox}[title={\small{\textbf{Challenge 2: Disparities between Concrete and Symbolic Execution}}}, 
                  colback=gray!30!white,
                  colframe=black,
                  top=1pt,
                  bottom=1pt]
\small{Accurate extraction of crash constraints requires achieving execution path consistency across concrete and symbolic environments.}
\end{tcolorbox}

{
\begin{table*}[t]
\centering
\normalsize
\caption{
Example of constraint stitching between a target function and call edge.
}
\resizebox{\linewidth}{!}{
\begin{tabular}{l|c|c}
\toprule
\multirow{2}{*}{\begin{tabular}[c]{@{}c@{}}\textbf{Crash constraints} ($\scr$)\\ (\textbf{Source Function}:~\texttt{add\_elem})\\ (\textbf{Symbolic Value of Parameters}:\\~\texttt{data}: $S1$,~\texttt{data->size}: $S2$,\\~\texttt{buf}: $S3$,~\texttt{size}: $S4$)\end{tabular}} & \multicolumn{2}{c}{\begin{tabular}[c]{@{}c@{}}\small{\textbf{Stitched Constraints} ($\pcr \land (\scr\odot{}a^{f}_{i})$) \ \ \ \  \  \ \textbf{Caller Function}:~\texttt{process\_req}}\\ (\textbf{Symbolic Value of Parameters}:~\texttt{r}: $S5$,~\texttt{idx}: $S6$,~\texttt{buf}: $S7$,~\texttt{sz}: $S8$)\end{tabular}}                                                                                                                                                                          \\ \cline{2-3} 
                                                                                                                                                                     & \begin{tabular}[c]{@{}c@{}}\textbf{Call Site 14}\\ (\textbf{$\pcr_{14}$}:~\texttt{(S6<POOL\_SIZE)}$\land$\texttt{(S5==UPDATE)},\\~\textbf{Argument values} ($a^{f}_{14}$)~\textbf{ to }~\texttt{add\_elem}: \\\{\texttt{data}: \texttt{const\_glob}, \texttt{data->size} = $S8$, \\ \texttt{buf}: $S7$, \texttt{size} = $S8$\})\end{tabular} & \begin{tabular}[c]{@{}c@{}}\textbf{Call Site 20}\\ (\textbf{$\pcr_{20}$}: ~\texttt{(S6<POOL\_SIZE)}$\land$\texttt{(S5==NEW)},\\~\textbf{Argument values} ($a^{f}_{20}$)~\textbf{ to }~\texttt{add\_elem}:\\\{\texttt{data}: \texttt{const\_glob}, \texttt{data->size} = $S8$, \\ \texttt{buf}: \texttt{"HDR"}, \texttt{size} = 3\})\end{tabular} \\ 
                                                                                                                                                                     \midrule
     \encircle{$c_{1}$}~\texttt{(S1==NULL)}                                                                                                                                                                & \begin{tabular}[c]{@{}c@{}}\texttt{(S6<POOL\_SIZE)}$\land$\texttt{(S5==UPDATE)}\\ $\land$\colorbox{pink}{\texttt{(const\_glob==NULL)}}~\textcolor{red}{\faRemove}\end{tabular}                                                                                             & \begin{tabular}[c]{@{}c@{}}\texttt{(S6<POOL\_SIZE)}$\land$\texttt{(S5==NEW)}\\ $\land$\colorbox{pink}{\texttt{(const\_glob==NULL)}}~\textcolor{red}{\faRemove}\end{tabular}                                                                                               \\ %
     
    \rowcolor{black!15} \encircle{$c_{2}$}~\texttt{(S1!=NULL)}$\land$\texttt{(S2<S4)}                                                                                                                                                               & \begin{tabular}[c]{@{}c@{}}\texttt{(S6<POOL\_SIZE)}$\land$\texttt{(S5==UPDATE)}\\ $\land$\texttt{(const\_glob!=NULL)}$\land$\colorbox{pink}{\texttt{(S8<S8)}}~\textcolor{red}{\faRemove}\end{tabular}                                                                                                 & \begin{tabular}[c]{@{}c@{}}\texttt{(S6<POOL\_SIZE)}$\land$\texttt{(S5==NEW)}\\ $\land$\texttt{(const\_glob!=NULL)}$\land$\texttt{(S8<3)}~\textcolor{green}{\faCheck}\end{tabular}                                                                                               \\ %
      
     \encircle{$c_{3}$}~\texttt{(S1!=NULL)}$\land$\texttt{(S3==NULL)}                                                                                                                                                                & \begin{tabular}[c]{@{}c@{}}\texttt{(S6<POOL\_SIZE)}$\land$\texttt{(S5==UPDATE)}\\ $\land$\texttt{(const\_glob!=NULL)}$\land$\texttt{(S7==NULL)}~\textcolor{green}{\faCheck}\end{tabular}                                                                                                 & \begin{tabular}[c]{@{}c@{}}\texttt{(S6<POOL\_SIZE)}$\land$\texttt{(S5==NEW)}\\ $\land$\texttt{(const\_glob!=NULL)}$\land$\colorbox{pink}{\texttt{("HDR"==NULL)}}~\textcolor{red}{\faRemove}\end{tabular}                                                                                               \\ %
     
     \rowcolor{black!15} \encircle{$c_{4}$}~\texttt{(S1!=NULL)}$\land$\texttt{(obj\_size(S3)<S4)}                                                                                                                                                                & \begin{tabular}[c]{@{}c@{}}\texttt{(S6<POOL\_SIZE)}$\land$\texttt{(S5==UPDATE)}\\ $\land$\texttt{(const\_glob!=NULL)}$\land$\texttt{(obj\_size(S7)<S8)}~\textcolor{green}{\faCheck}\end{tabular}                                                                                                 & \begin{tabular}[c]{@{}c@{}}\texttt{(S6<POOL\_SIZE)}$\land$\texttt{(S5==NEW)}\\ $\land$\texttt{(const\_glob!=NULL)}$\land$\colorbox{pink}{\texttt{(obj\_size("HDR")<3)}}~\textcolor{red}{\faRemove}\end{tabular}                                                                                               \\ 
\bottomrule
\end{tabular}
}
\label{tbl:backtrack}
\end{table*}
}

\subsection{Backtracking and Crash Validation}
\label{overview:phasethree}

To validate the potential vulnerabilities identified in function $f$, we must examine the call graph to identify all execution paths leading to $f$.
For our running example, we identify the call edges in program $P$ as $e1$=\code{(process_req, add_elem)} and $e2$=\code{(handle_req, process_req)}.
To verify the feasibility of the crashes in $f$, we need to perform \emph{Constraint Stitching} of the \emph{Crash Constraints} of $f$ with the \emph{Edge Constraints} of $e1$ and $e2$.

Beginning with the first call edge $e1$, we generate a specialized driver program $\dr^{e1}$ (with $pf$ as the target function) following the steps described in \sect{overview:phaseone}.
However, since we are interested only in \emph{Edge Crashes}
(in this case, there are two callsites for \code{add_elem}). 
Similar to \sect{overview:phasetwo}, we use $\dr^{e1}$ to collect the \emph{Edge Constraints} for each call site.
The second and third columns of \tbl{tbl:backtrack} show the two call sites to \code{add_elem} in \code{process_req} along with the path constraints and symbolic values of arguments passed to \code{add_elem}.
Finally, we stitch the \emph{Crash Constraints} of $f$ with the \emph{Edge Constraints} of $e1$ to check if the crash is feasible from the call site.
The cells in the $2^{nd}$ and $3^{rd}$ column of~\tbl{tbl:backtrack} show the stitched constraints at corresponding call-sites of~\code{process_req} for each of the crashes in~\code{add_elem}.
The satisfiable (\textcolor{green}{\faCheck}) and unsatisfiable (\textcolor{red}{\faRemove}) constraints are marked. In addition, for unsatisfiable constraints, the corresponding unsatisfiable component is \colorbox{pink}{highlighted}.
The satisfiability of the stitched constraint indicates whether the corresponding crash is feasible from the call-site or not.

If this stitching is satisfiable, we proceed to the next level in the call graph, now using the stitched constraint unto its caller.
The next call edge is $e2$=\code{(handle_req, process_req)}.
This is shown in~\tbl{tbl:backtrack2}, where the feasible constraints in~\code{process_req} (\tbl{tbl:backtrack}) are backtracked unto~\code{handle_req} using the new (stitched) constraints.
We follow the same procedure in~\code{handle_req} and discard infeasible crashes.
We attempt to stitch all the feasible crashes until either we reach~\code{main} or the backtracking fails (see \sect{partial_backtrack} for more details).
Finally, all the crashes feasible from~\code{main} are considered as true crashes and are reported to the developer.

{
\begin{table*}[]
\centering
\caption{Example of backtracking between two call edges.
}
\scriptsize
\begin{tabular}{l|c}
\toprule
\multirow{2}{*}{\begin{tabular}[c]{@{}c@{}}\textbf{Crash constraints} ($\scr$)\\ (\textbf{Source Function}:~\texttt{process\_req})\\ (\textbf{Symbolic Value of Parameters}:\\~\texttt{r}: $S5$,~\texttt{idx}: $S6$,~\texttt{buf}: $S7$,\\~\texttt{sz}: $S8$)\end{tabular}} &  \begin{tabular}[c]{@{}c@{}}\textbf{Stitched Constraints} ($\pcr \land (\scr\odot{}a^{f}_{i})$) \ \ \ \ \ \textbf{Caller Function}:~\texttt{handle\_req}\\ (\textbf{Symbolic Value of Parameters}:~\texttt{r}: $S9$,~\texttt{gidx}: $S10$,~\texttt{sockfd}: $S11$) \\
(\textbf{Fresh symbolic values}:$S12$)\end{tabular} \\ \cline{2-2} 
                  &  \begin{tabular}[c]{@{}c@{}}\textbf{Call Site 14} : \ \ \ \ \ \ (\textbf{$\pcr_{14}$}:~\texttt{(S9==NEW)}$\land$\texttt{(S12>0)},\\~\textbf{Argument values} ($a^{f}_{14}$)~\textbf{ to }~\texttt{process\_req}: \\\{\texttt{r}: $S9$, \texttt{idx}: $S10$, \texttt{buf}: \texttt{stack\_obj}, \texttt{size}:$S12$\})\end{tabular} \\
                  \midrule
  \begin{tabular}[c]{@{}c@{}}\encircle{$c_{2}$}~\texttt{(S6<POOL\_SIZE)}$\land$\texttt{(S5==NEW)}\\ $\land$\texttt{(const\_glob!=NULL)}$\land$\texttt{(S8<3)}\end{tabular}                                                                                                                                                               &    \begin{tabular}[c]{@{}c@{}}\texttt{(S9==NEW)}$\land$\texttt{(S12>0)}$\land$\texttt{(S10<POOL\_SIZE)}$\land$\texttt{(S9==NEW)}\\ $\land$\texttt{(const\_glob!=NULL)}$\land$\texttt{(S12<3)}~\textcolor{green}{\faCheck}\end{tabular} \\
  
  \rowcolor{black!15} \begin{tabular}[c]{@{}c@{}}\encircle{$c_{3}$}~\texttt{(S6<POOL\_SIZE)}$\land$\texttt{(S5==UPDATE)}\\ $\land$\texttt{(const\_glob!=NULL)}$\land$\texttt{(S7==NULL)}\end{tabular}                                                                                                                                                               &    \begin{tabular}[c]{@{}c@{}}\texttt{(S9==NEW)}$\land$\texttt{(S12>0)}$\land$\texttt{(S10<POOL\_SIZE)}$\land$\colorbox{pink}{\texttt{(S5==UPDATE)}}\\ $\land$\texttt{(const\_glob!=NULL)}$\land$\colorbox{pink}{\texttt{(stack\_obj==NULL)}}~\textcolor{red}{\faRemove}\end{tabular} \\
  
  \begin{tabular}[c]{@{}l@{}}\encircle{$c_{4}$}~\texttt{(S6<POOL\_SIZE)}$\land$\texttt{(S9==UPDATE)}\\ $\land$\texttt{(const\_glob!=NULL)}$\land$\texttt{(obj\_size(S7)<S8)}\end{tabular}                                                                                                                                                               &   \begin{tabular}[c]{@{}c@{}}\texttt{(S9==NEW)}$\land$\texttt{(S12>0)}$\land$\texttt{(S10<POOL\_SIZE)}$\land$\colorbox{pink}{\texttt{(S9==UPDATE)}}\\ $\land$\texttt{(const\_glob!=NULL)}$\land$\texttt{(obj\_size(stack\_obj)<S12)}~\textcolor{red}{\faRemove}\end{tabular} \\
\bottomrule
\end{tabular}
\label{tbl:backtrack2}
\end{table*}
}

The constraint stitching process presents significant technical challenges when dealing with complex types.
Most state-of-the-art symbolic execution engines, such as KLEE~\cite{klee_repo} and angr~\cite{cheng2016binary}, cannot handle symbolic sizes for arrays.
Another limitation is that most symbolic execution engines cannot handle symbolic pointers; they require concrete memory addresses.
To work around this, we allocate concrete memory in the harness and make only the values within these memory regions symbolic.
For example, in \lst{lst:runexample} (a), we create a concrete pointer \code{data} but make each field of \code{struct pool} symbolic inside it.
This approach creates challenges for tracking null pointer conditions - when a null dereference crash occurs at Line 7, the symbolic engine detects the crash but cannot express it as a symbolic constraint since the null pointer is a concrete value, not a symbolic one. 
Finally, we also need to ensure that the stitching process removes any constraints that are introduced by the harness.

\begin{tcolorbox}[title={\small{\textbf{Challenge 3: Stitching Complex Types}}}, 
                  colback=gray!30!white,
                  colframe=black,
                  top=1pt,
                  bottom=1pt]
\small{To accurately stitch constraints, we need to ensure that we can stitch 
(1) \emph{Arrays and symbolic sizes,}
(2) \emph{Pointers,}and 
(3) \emph{Remove Harness-induced constraints.}
}
\end{tcolorbox}

\section{Griller: Design of a Reactive BOUT System}

\begin{figure*}[t]
\centering
\includegraphics[scale=0.61]{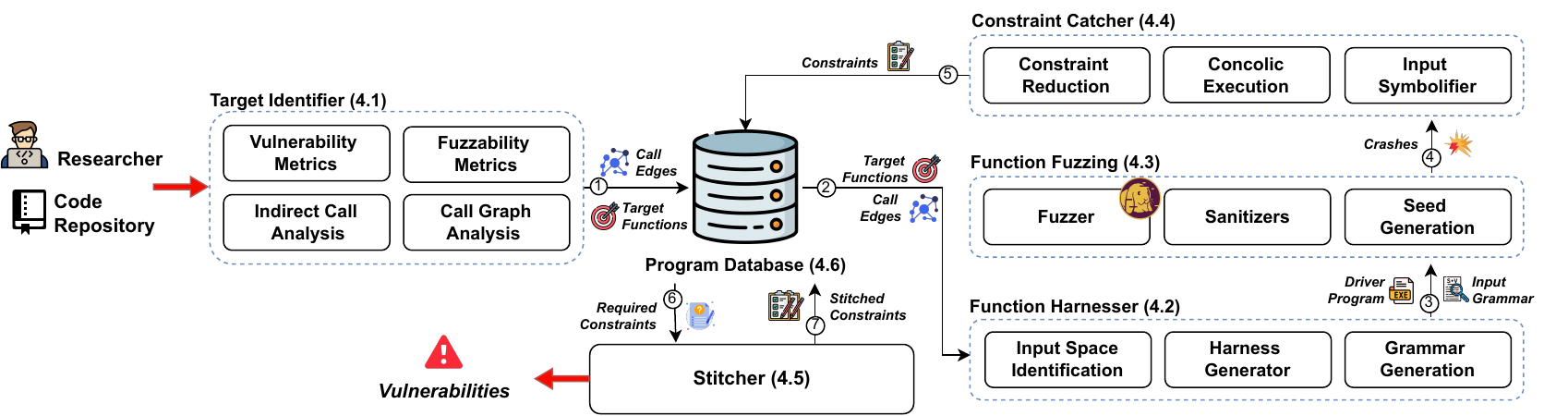}
\caption{Design Overview of \ourtool{}}
\label{fig:workflow}
\end{figure*}

In this section, we describe the prototype implementation of the~\ourtool{} Framework and discuss the design of the various components.
The design of \ourtool{} follows a staged approach, as illustrated in \fig{fig:workflow}. 
\ourtool{} consists of five major components (1) Target Identifier, (2) Stitcher, (3) Function Harnesser, (4) Function Fuzzer, (5) Constraint Catcher.
Our system is designed to be staged, where each component executes in isolation and stores its result in the \emph{Program Database}.
We will discuss each component in detail.
Our implementation details are discussed in detail in the \apdx{apdx:implementation}

\subsection{Target Identifier}
\label{sec:targetidentifier}

The \emph{Target Identifier} is responsible for identifying functions within the target application that are suitable for function-level testing, as well as determining the call edges necessary to validate crashes observed in these target functions.
While \ourtool{} allows researchers to manually specify functions of interest, it also incorporates a set of heuristics to systematically identify and prioritize target functions.
To facilitate the identification of target functions, \ourtool{} employs two primary heuristics: the \emph{Vulnerability Metric} and the \emph{Fuzzability Metric}.

\subsubsection{Vulnerability Metric}
\label{designone:vulnmetrics}
The Vulnerability metric is meant to assess the likelihood of a function containing vulnerabilities.
We used the Vulnerability metric proposed by LEOPARD~\cite{leapord} as the baseline for our heuristics.

\subsubsection{Fuzzability Metric}
\label{designone:fuzzmetrics}
The Fuzzability metric is meant to assess the ease of fuzzing a function.
Functions with complex/larger parameter types can be relatively harder to fuzz, since the input space is larger and decreases fuzzing efficiency.
We compute a Type Score for each function that estimates the complexity of generating its input space.
We use these metrics to classify each function into one of three priority levels - HIGH, MEDIUM or LOW.
We provide a detailed description of these metrics and our implementation in Appendix~\ref{apdx:metrics}.

\subsubsection{Call Graph Analysis}
\label{designone:callgraph}
To identify the call edges needed to validate crashes, \ourtool{} enumerates caller\textrightarrow callee paths in the program's directed call graph that reach the target function.
We treat cycles (including direct and indirect recursion) uniformly by visiting each call edge at most once per path; if we encounter direct recursion, we include at most one self-edge.

\subsubsection{Indirect Call Analysis}
\label{designone:indirectcall}
Indirect calls and function pointers are prevalent in C programs~\cite{indirectCalls}, making it essential to handle them correctly to obtain an accurate call graph.
While this is a well-explored research area~\cite{li2025redefining, mlta-ccs19}, we opted for a simpler solution.
We identify potential target functions for each indirect call based on two criteria: (1) the function must have had its address captured somewhere in the program, and (2) the function signature must match the call site signature.
Using these identified functions, we transform indirect calls into direct calls at their respective call sites, which significantly simplifies subsequent analysis components.

\subsection{Function Harnesser}
\label{sec:funcharnesser}

The \emph{Function Harnesser} generates driver programs for both target functions and call edges retrieved from the Program Database.
First, it identifies the variables that make up the input space of the target function using lightweight static analysis.
Then, it creates a harness with appropriate input generators tailored to each identified variable's type.
For call-edge testing specifically, we apply optimizations such as path pruning to improve efficiency.

\subsubsection{Input Space Identification}
For a target function $f$, its input space consists of (1) variables (function arguments and accessed global variables) and (2) external sources  (e.g., stdin, environment variables).
We focus on (1) here; external sources are addressed via runtime hooks(\sect{sec:funcfuzz}).
The variables can be of complex types containing nested pointers to other types.
Our analysis recursively traverses each type to identify all the variables and their corresponding types to generate the input space.
As highlighted in \emph{Challenge 1}, our harness generation must satisfy three key properties: completeness, type and context awareness, and minimality. 
We address each of these properties below

\begin{itemize}
\item \textbf{Completeness:} 
We track cast operations on variables in the input space to determine their runtime destination types.
Our analysis follows data flow through the function to identify all cast sites and capture the types.
\item \textbf{Type and Context Awareness:} 
We employ CCured~\cite{necula2005ccured}-based type inference to distinguish array pointers from single objects by analyzing pointer arithmetic and indexing operations. 
Our Locksmith~\cite{pratikakis2011locksmith}-inspired co-relational analysis tracks data flow between parameters to establish array-size associations.
\item \textbf{Minimality:} 
We construct a dominator tree for the target function to identify def-dominating variables which need not be generated as inputs. 
For structures, we perform field-sensitive analysis to track which members are accessed by the function and its callees, excluding unaccessed fields from generation.
\end{itemize}

\subsubsection{Harness Generation}
\label{sec:harnessgen}

Algorithm~\ref{algo:typegen}(in Appendix) presents our harness generation approach, which systematically constructs input generators based on their types.
Base types (\code{int}, \code{float}, etc.) are handled directly by reading the appropriate number of bytes from \code{stdin}.
Composite types require a more sophisticated approach due to their potential nesting complexity, such as structures containing arrays of structures with pointer fields or multi-dimensional arrays.

Our algorithm employs a two-phase strategy for composite types: first generating the necessary control structures, then processing nested elements.
For structures, we allocate memory based on the structure's size, then invoke generators for each accessed field.
Array generation involves determining the size and generating each element through the appropriate type generator.
Special cases including function pointers, opaque types are detailed in Appendix~\ref{appendix:opaque_types}.

During harness generation, we also keep track of all the processed types and use that information to generate an \emph{input grammar} that specifies how data from standard input is structured to create the function's input space.
\lst{lst:grammar} shows the input grammar for the \code{add_elem} function from \lst{lst:runexample}.

\begin{listing}
\begin{minted}[breaklines,fontsize=\scriptsize]{c}
// grammar for the target function (add_elem)
<start> ::= <arg_0_data> <arg_2_size> <arg_1_buf> 
// Argument 1: data, type : struct pool * (was void *)
<arg_0_data> ::= structptr[<5:avail> <7:size> <6:buf>] 
// Arguments 2 & 3: buf and size, size is correlated to buf
<arg_1_buf> ::= array[int8, <arg_2_size>]
<arg_2_size> ::= int32
// grammar for the elements of struct poolß
<5:avail> ::= int8
<7:size> ::= int32
<6:buf> ::= array[int8, <7:size>]
\end{minted}
\caption{
Input grammar generated for the trampoline program for \code{add_elem} in \lst{lst:runexample}.
}
\label{lst:grammar}
\end{listing}

\subsection{Function Fuzzer}
\label{sec:funcfuzz}

The Function Fuzzer systematically tests the generated driver program to either discover vulnerabilities in target functions (and their callees) or find inputs reaching specific call sites when testing call edges.
Effective fuzzing requires four essential components: 
(1) a fuzzing harness that interfaces with the target code, 
(2) high-quality seed inputs, 
(3) sanitizers, and 
(4) a fuzzing engine. 
The driver program generated in the previous step provides our fuzzing harness.
The remaining three components are provided by the Function Fuzzer, which we describe in detail below.

\subsubsection{Seed Generation}
Research has consistently shown that high-quality seeds significantly improve fuzzing effectiveness~\cite{rebert2014optimizing, herrera2021seed, lyu2018smartseed}.
Fuzzers also need a seed corpus to start testing applications.
Since the driver program's input format differs from that of the original application, we need specialized seed generation techniques. 
\ourtool{} employs two complementary approaches for generating effective seed inputs: \textit{Grammar-Based} and \textit{Symbolic Execution-Based}.

\begin{itemize}
\item \textbf{Grammar-Based} : 
The primary seed generation method in \ourtool{} leverages the \texttt{input grammar} (\lst{lst:grammar}) to generate seed inputs.
Our grammar-based seed generator systematically creates well-formed inputs by varying fields within the constraints of the grammar. 
It produces diverse inputs by strategically setting some pointers to null, varying array sizes, and setting different value combinations for primitive types. 
\item \textbf{Symbolic Execution-Based} :
As a complementary approach, \ourtool{} can also generate seeds through symbolic execution of the target function. 
For this, we modify the driver program to create symbolic variables for each input rather than reading from stdin.
We then execute the function symbolically for a predetermined time period, collecting the explored states. 
We use the symbolic values generated for every variable to build seed inputs based on the structure defined in the \texttt{input grammar}.
\end{itemize}

\subsubsection{Sanitizers}
Sanitizers are an important part of the fuzzing ecosystem, identifying silent bugs that might corrupt the memory of the application without leading to crashes. 
However, as mentioned in \emph{Challenge 2}, \ourtool{} requires running the same inputs using pre-constrained symbolic execution in later stages.
To enforce a much stricter runtime policy, we modified ASan with additional checks to detect crashes at their root cause, such as the point where a pointer goes out of bounds due to incorrect arithmetic rather than at the point where the out-of-bounds pointer is used. 

\subsubsection{Fuzzer}
We employ AFL++~\cite{AFLplusplus-Woot20} as our primary fuzzing engine, with two key modifications for function-level testing.
First, to handle external input sources (\sect{sec:funcharnesser}), we implement runtime hooks that intercept library calls (\code{read}, \code{recv}, \code{getenv}) and redirect them to stdin, giving AFL++ complete control over all input channels (details in Appendix~\ref{apdx:runtimehooks}).
Second, since function-level fuzzing produces crashes far more frequently than whole-program fuzzing, we modified AFL++ to retain crashes with new coverage in its mutation queue rather than discarding them—enabling continued exploration beyond initial crash sites.
While our implementation uses AFL++, \ourtool{}'s modular design supports any coverage-guided fuzzer with minimal adaptation overhead.

\subsection{Constraint Catcher}
\label{sec:constraintgen}

The Constraint Catcher extracts symbolic constraints representing crash conditions through \emph{pre-constrained} symbolic execution (forcing the symbolic engine to follow the exact path taken by each crashing input).
We employ KLEE~\cite{klee_repo} for its tight LLVM integration and efficient IR-level symbolic execution capabilities.

\subsubsection{Input Symbolizer}
To ensure execution path parity between fuzzing and symbolic execution (\emph{Challenge 2}), it is essential that the exact same input values exercised during fuzzing are replayed during symbolic execution.
However, AFL++ and KLEE expect inputs in different formats: AFL++ consumes a raw byte stream from \texttt{stdin}, while KLEE requires a structured \texttt{KTEST} file, which encodes named symbolic variables and their concrete values.
The \emph{Input Symbolizer} bridges this gap by translating fuzzer-generated inputs into the \texttt{KTEST} format required by KLEE.
It leverages the \texttt{input grammar} (\lst{lst:grammar}) to parse the raw byte stream and map each segment to its corresponding symbolic variable. 
The Input Symbolizer reconstructs the expected variable layout, extracts values for each symbolic variable, and serializes them into the \texttt{KTEST} format with correct naming and byte sequences.

\subsubsection{Concolic Execution}
\label{sec:concolic}

We provide KLEE with the input in the \texttt{KTEST} format to enable pre-constraint symbolic execution.
While KLEE automatically collects path constraints when we run \emph{pre-constraint} symbolic execution with the crashing input.
It has no way of capturing the root-cause constraints for the crash occurring at the callsite.
To extract these, we instrument the driver code by inserting assertions at crash locations identified during fuzzing. 
As shown in \lst{lst:constraints}, we implement multiple types of safety checks, \code{grill_check} functions detect null pointer dereferences, while \code{grill_check_buf} verifies buffer size constraints.
Our implementation inserts these assertions directly into the LLVM-IR at memory-operation related instructions like LOAD, STORE, and GEP. 
We also add hooks for common standard library functions such as \code{memcpy} and \code{strcpy} that frequently cause memory safety violations, since external library functions are not executed symbolically by KLEE. 
Additionally, we've created a YAML-based configuration system that allows for easily adding new function hooks.

To accurately capture \emph{Edge Constraints}, we must record the symbolic values of each argument at the callsite. 
This enables precise constraint stitching between parent and child functions. 
To achieve this, we developed a \emph{TypePrinter} component, which systematically traverses each argument at the callsite and serializes its symbolic value.
We provide more details of this in \apdx{apdx:typeprinter}.

\subsubsection{Constraint Reduction}
As discussed in \emph{Challenge 3}, we must augment KLEE to accurately capture array sizes and null pointers, since these are critical for precise vulnerability analysis.
We introduce additional symbolic variables to track array sizes dynamically.
We add a runtime tracking mechanism that records pointers with their corresponding symbolic sizes and use this to reason about the sizes of the pointers during constraint generation.
For pointer nullness, we also create additional symbolic variables to encode it explicitly (e.g., a boolean variable indicating whether each pointer is NULL).
These additions enable us to express constraints like "buffer size must equal the size parameter" or "pointer must be non-NULL at dereference," which standard symbolic execution cannot capture directly.

However, our instrumentation also introduces spurious constraints as a side effect.
Array iteration loops in the harness generate unrelated size constraints.
Similarly, our nullness encoding produces artificial null-checks that don't exist in the original code.
To address this, we perform \emph{constraint reduction} that systematically identifies harness-induced constraints and eliminates them from the final constraint set.
This ensures our analysis captures only the essential conditions for each crash.
Details of our tracking mechanisms and reduction algorithms are provided in ~\apdx{apdx:reduction}.

\begin{listing}
\begin{minted}[linenos, xleftmargin=18pt, breaklines,fontsize=\scriptsize,escapeinside=||]{c}
int add_elem(void *data, 
             const char *buf, 
             unsigned size) {
  struct pool *p = (struct pool *)data;
|\text{\faArrowRight}| grill_check(p != null);
  if (p->avail) { 
   p->buf = malloc(p->size*sizeof(char));
|\textcolor{green}{\faArrowRight}|   grill_track_size(p->buf, p->size * sizeof(char));
   if (p->buf) {
     p->avail = 0;
|\text{\faArrowRight}|   grill_check(p->buf != null);
|\text{\faArrowRight}|   grill_check_buf(p->buf, size);
|\text{\faArrowRight}|   grill_check(buf != null);
|\text{\faArrowRight}|   grill_check_buf(buf, size);
     memcpy(p->buf, buf, size);
   } ... }

grill_check_buf(void *buf, unsigned bufsize) {
  klee_assert(grill_get_tracked_size(buf) >= bufsize);
}

\end{minted}
\caption{Snippet showing Root Cause Assertions added to the function \code{add_elem} in \lst{lst:runexample}.~\text{\faArrowRight}~indicates the assertions added.~\textcolor{green}{\faArrowRight}~indicates the runtime hooks added.}
\label{lst:constraints}
\end{listing}

\subsection{Stitcher}
\label{sec:stitcher}

The primary goal of the stitcher is to combine the \emph{Crash Constraints} (path constraints and root cause constraints) with the \emph{Edge Constraints} (path constraints and symbolic argument values) for all possible paths to reaching the target function and check for their satisfiability.
To understand what types need to be stitched, \ourtool{} captures the mapping between the arguments of the parent and the child functions at the callsite.

Stitching types involves generating constraints that ensure the values passed from a parent function to a child function are consistent across the call edge. 
For basic types (such as integers or chars), we represent each value as a bitvector and generate equality constraints for each segment.
If the types differ in size, we perform bit extension or truncation as needed to align their sizes before generating the equality constraint.
For arrays, stitching requires handling both the array size and the individual elements.
We generate a constraint equating the array sizes between caller and callee, and then, for each index up to the minimum of the two sizes, we generate equality constraints for the corresponding elements. 
For each pointer-type argument, we also stitch the symbolic variable encoding the nullness of the pointer.

\subsubsection{Partial Backtracking}
\label{partial_backtrack}
There might also be cases where we were unable to fully explore and generate constraints for a call-edge. 
This is mainly due to limitations of fuzzing.
However, even in this scenario, ~\ourtool{} can reason about the vulnerability by backtracking it until the failing call-edge.
This is referred to as \emph{partial backtracking}.
Since partial backtracking lacks the complete program context necessary for   validation, reported bugs may be false positives.
To help distribute effort, \ourtool{} distinguishes between partial and complete backtracking results in its output reporting.

\subsection{Program Database}
\label{sec:constraintdb}
The \emph{Program Database} serves as the central repository in \ourtool{}, enabling a staged and modular workflow. 
It stores all relevant information about the target program, including identified target functions, call edges, generated driver programs, discovered crashes, extracted constraints, and symbolic argument values. 
Each component of \ourtool{} interacts with the Program Database to fetch required data and persist results.
This on-demand execution avoids redundant computation and ensures that each function or call edge is analyzed only as needed.
By centralizing all intermediate and final analysis results, the Program Database also enables efficient incremental analysis.

\section{Evaluation}
\label{sec:evaluation}

We conduct a comprehensive evaluation of \ourtool{} on real-world applications and libraries to assess its effectiveness and performance. Our evaluation aims to assess the following research questions:

\begin{itemize}[leftmargin=*]
    \item \textbf{RQ1: Known Vulnerability Detection.} How effective is \ourtool{} in finding known vulnerabilities?
    \item \textbf{RQ2: Effectiveness of Components.} How effective is each component of the \ourtool{} framework?
    \item \textbf{RQ3: New Vulnerabilities Detection.} How effective is \ourtool{} in discovering new vulnerabilities?
\end{itemize}

% \vspace{1em}
\indent\textbf{Setup.}
\label{subsect:setup}
We performed all experiments on a VM (n2-standard-128) on the Google Cloud platform with 128 cores and 512GB RAM. 
All fuzzing runs were bound to a single core.

\subsection{RQ1: Finding Known Vulnerabilities}
\label{sec:rq1}

In this experiment we evaluate the performance of \ourtool{} in identifying vulnerabilities in a controlled setting.
We compare \ourtool{}'s performance on this task to the performance of three state-of-the-art methods: AFGen~\cite{AFGen}, AFL++~\cite{AFLplusplus-Woot20}, and Beacon~\cite{9833751}. 
In this experimetn we do not run these tools ourselves, but rely on the published results from the AFGen paper.
We give a detailed comparison to AFGen because it has been extensively compared to other function-level fuzzing tools, as well as to directed and undirected fuzzers.
Because AFGen source code is not available, we cannot perform a direct comparative implementation evaluation. 
Instead, we use the same dataset as AFGen and compare our results to their published findings.

\vspace{0.3em}
\indent\textbf{Benchmark.} 
Our evaluation uses a refined subset of the dataset from AFGen~\cite{AFGen}. 
Specifically, we selected actively maintained projects, vulnerabilities that have been verified and reproducible, \ie{} have crashing input.
We provide a detailed explanation of our dataset refinement methodology, including selection criteria and excluded targets, in Appendix~\ref{appendix:dataset-filtering}.
\tbl{tab:dataset_breakdown} summarizes our dataset along with the set of 48 known vulnerabilities (CVEs), Fuzzing Configurations (Versions / Programs), unique target functions, and unique call edges.

\begin{table}[]
\caption{Summary of Dataset}
\label{tab:dataset_breakdown}
\resizebox{\columnwidth}{!}{
\begin{tabular}{c|c|c|c|c}
\toprule
\textbf{Target} & \textbf{\begin{tabular}[c]{@{}c@{}}Config\\ Ver / Prog\end{tabular}} & \textbf{CVEs} & \textbf{\begin{tabular}[c]{@{}c@{}}Unique\\ Target Functions\end{tabular}} & \textbf{\begin{tabular}[c]{@{}c@{}}Unique\\ Call Edges\end{tabular}} \\ \hline
ngiflib         & 3 / 1                                                           & 9                        &         7                                                             &        15                                                              \\ \hline
ffjpeg          & 2 / 1                                                           & 12                       &    9                                                                  &          9                                                            \\ \hline
tcpreplay       & 4 / 5                                                           & 16                       &         14                                                             &     46                                                                 \\ \hline
jasper          & 1 / 1                                                           & 8                        &            7                                                          & 19                                                                     \\ \hline
lupng           & 1 / 1                                                           & 3                        &                       2                                               &    7                                                                  \\ \midrule
\textbf{Total} & 11 / 9  & 48 & 39 & 96 \\
\bottomrule
\end{tabular}
}
\end{table}

\vspace{0.3em}
\indent\textbf{Method.} 
\ourtool{} is provided with the list of target functions (\ie{} function containing the vulnerability) by manually analyzing the reports for each vulnerability.
Our time parameters were selected as follows. 
Baseline tools analyze complete applications and use an established 24-hour per-target evaluation budget. 
\ourtool{} operates at the function and call-edge granularity, so we did a best-effort attempt provide equivalent computational resources.
We performed fuzzing (for both target functions and call edges) for a time frame of 30 minutes to 3 hours, depending upon the complexity of the function.
We only used randomized grammar-based seed generation, to be consistent with AFGen.
For constraint extraction runs, we allocated \setupConstraintRun{} each, since most of them finished within 30 mins.
We did not impose an overall time budget to ensure that all valid crashing inputs could be properly converted into constraints.
Similarly, we allocated a timeout of \setupStitchingTime{} for each constraint being stitched, but didn't impose an overall time budget to ensure that all constraints could be stitched.
In our experience, most of the constraint stitching finished within 1 minute.
We manually verified each of the stitching results to ensure that they were related to the vulnerability.

\begin{table}[h]
\caption{
Summary of Real-world vulnerabilities reproduced by \ourtool{} and AFGen.}
\scriptsize
\label{tab:rq1comparison}
\centering

\begin{tabular}{c|c|c|c|c|c}
\toprule
\textbf{Target} & \textbf{Vulnerabilities} & \textbf{Griller} & \textbf{AFGen} & \textbf{AFL++} & \textbf{Beacon} \\ \hline
ngiflib         & 9                        & 9                 & 6         &  6  & 7   \\ \hline
ffjpeg          & 12                       & 7               & 9        & 8   & 8     \\ \hline
tcpreplay       & 16                       & 7                & 10       & 1   & 0   \\ \hline
jasper          & 8                        & 3                 & 5         & 0  & 0    \\ \hline
lupng           & 3                        & 2                & 3      & 1    & 0     \\ \midrule
\textbf{Total} & 48 & 28 & 33 & 16 & 15\\ \bottomrule
\end{tabular}
\end{table}

% \vspace{1em}
\indent\textbf{Results.} 
\tbl{tab:rq1comparison} presents a comparative analysis of vulnerabilities detected by each tool across different targets.
\ourtool{} identified \rqtwoTPCount{} true positives and reported \rqtwoFPCount{} false positives.
\ourtool{} outperformed AFL++ and Beacon by a significant margin and found a comparable number of bugs as AFGen.
\emph{Notably, \ourtool{} successfully identified 4 vulnerabilities that AFGen failed to detect, demonstrating complementary detection capabilities}; one such case is illustrated by \lst{lst:ngiflib_tp} (in Appendix).
This is mainly due to our approach and superior handling of null pointers as described in \apdx{apdx:symbolic_sizes}.

\textit{Backtracking:}
Our backtracking mechanism successfully traced 15 vulnerabilities completely to the main function, with all of these cases confirmed as true positives (100\% precision).
For the remaining cases, backtracking was only partially successful, stopping at the highest reachable program point before main. 
This partial backtracking limitation was the primary source of our false positives, as incomplete path reconstruction prevented full vulnerability validation.
All 6 false positives are valid vulnerabilities up to the program point we successfully backtracked to, but become invalid when considering the complete call graph context.
However, it is important to note that \ourtool{} reports partial and complete backtracking cases separately.

\textit{Triggering Vulnerabilities:}
While fuzzing the target functions to find vulnerabilities, \ourtool{} was able to trigger \rqtwoTotalUniqueLeafCrashes{} (77\%) vulnerabilities, out of \rqtwoTotalRequiredCrashes{} vulnerabilities.
\ourtool{} was able to find a crash in most functions within an average of 2 seconds.
\ourtool{} also found 6 crashes that no other fuzzer was able to trigger.
\emph{This demonstrates Griller's superior performance in function-level vulnerability identification compared to existing approaches.}
However, 9 of these triggered vulnerabilities could not be fully verified due to backtracking failures, which explains the reduced number of confirmed true positives in our final results.
\tbl{tab:rq2_leaf_effectiveness} (in Appendix) provides a break down of the crashes found per target. 

\textit{Performance vs AFGen:}
AFGen identified 5 additional vulnerabilities compared to \ourtool{} in our evaluation.
This difference can be attributed to the dataset composition, which consists of file-parsing applications.
\ourtool{} faces reachability challenges while backtracking functions with low depth (\ie{} top-level functions such as main).
This is because these functions have complex input requirements such as valid \textit{jpeg} or \textit{png} inputs.
However, it is hard for a fuzzer to generate valid file formats without seeds.
The unavailability of AFGen's source code prevented us from conducting comparative analysis on other application domains to assess relative performance across different vulnerability patterns.

\textit{Call Edge Performance:}
\tbl{tab:rq2_call_edges} (in Appendix) summarizes the effectiveness of \ourtool{} in reaching each callsite and the average number of unique paths to each callsite.
\ourtool{} was able to successfully reach \rqtwoCallEdgesReached{} non-unique call edges, which is \rqtwoCallEdgesReachedPercentage{} of the total call edges.
Of the \rqtwoCallEdgesReached{} call edges reached by fuzzing, \ourtool{} was able to generate constraints for \rqtwoCallEdgesConstraints{} (\rqtwoCallEdgesConstraintsPercentage{}) of them.

\textit{Failure Analysis:}
For each of the \rqtwoFNCount{}, we perform an analysis into why we were not able to find these crashes.
Our analysis revealed that \ourtool{} failed to trigger the root cause crash for 11 (22\%) CVEs. 
The remaining could not be stitched due to failures in generating valid inputs for the call edges.
We discuss this in more detail in \apdx{appendix:failureanalysis}.

\begin{tcolorboxfloat}[htbp]
\noindent\fcolorbox{black}{green!30!white}{
  \parbox{\dimexpr\linewidth-3\fboxsep-2\fboxrule\relax}{\parskip=1pt\parindent=0pt
RQ1 results show that \ourtool{} is capable of using Reactive \bout{} to find vulnerabilities with comparable performance to state-of-the-art and achieves 100\% precision with complete backtracking. 
}
}
\end{tcolorboxfloat}

\subsection{RQ2: Effectiveness of Components}
\label{sec:rq2}

In this research question, we assess the effectiveness of each of our components using the same dataset and setup as described in \sect{sec:rq1}.
For each component, we first explain the evaluation method and then present the corresponding results.

\subsubsection{Target Function Identification}
As mentioned in \sect{overview:phaseone}, this component identifies risky functions (\ie{} those containing vulnerabilities), and identifies the call-edges that most likely would trigger vulnerabilities in the identified risky functions.

\vspace{0.3em}
\indent\textbf{Method.}
We check how many of the target functions in our dataset (\tbl{tab:dataset_breakdown}) are successfully identified by our technique.
We also measure the percentage of non-unique call edges (\tbl{tab:dataset_breakdown}) that were successfully identified and the contribution of our indirect call promotion.

\vspace{0.3em}
\indent\textbf{Results.}
Among the known vulnerable functions, \rqoneMatchedTargetFunctionsHigh{} were identified by our filtering approach as HIGH priority, \rqoneMatchedTargetFunctionsMed{} as MEDIUM priority, and \rqoneMatchedTargetFunctionsLow{} as LOW priority.
Of the \datasetTotalCallEdges{} call edges in our dataset, \rqoneMatchedCallEdges{} were successfully matched with call edges from the vulnerability reports, corresponding to \rqoneMatchedCallEdgesPercentage{} of the total.

\emph{Indirect Call Promotion:}
To evaluate the effectiveness of indirect call promotion, we compared call edge matching with and without this technique.
Without indirect call promotion, only \rqoneMatchedCallEdgesWithoutPromotion{} call edges were successfully matched, corresponding to \rqoneMatchedCallEdgesWithoutPromotionPercentage{} of the dataset's call edges.
The indirect call promotion technique proved useful for larger programs.

\subsubsection{Harness Generation}
As mentioned in \sect{sec:harnessgen}, this component identifies the types of arguments and global variables accessed by a given function and generates a harness to generate data of the appropriate type and invoke the function.

\vspace{0.3em}
\indent\textbf{Method.}
We measure the effectiveness of type identification by the diversity of data types and their accuracy.
We also measure the number of functions for which harness generation was successful, and the diversity of corresponding harnesses.

\vspace{0.3em}
\indent\textbf{Results.}
\tbl{tab:commonTypes} (in Appendix) shows the diversity of the types identified by our type identification technique.
We were able to identify 6862 pointers as array types, of which 412 were assigned size relationships, demonstrating the effectiveness of our type inference and correlation analysis.

We verified the accuracy of the types through random sampling.
Specifically, we randomly sampled 20 functions and manually verified the accuracy of the identified types.
Our verification confirmed that \ourtool{} is able to generate harnesses for all the functions and properly generated appropriate types for all functions' parameters.

\subsubsection{Crash Constraint Generation}
As mentioned in \sect{sec:constraintgen}, this component identifies crashing constraint, \ie{} symbolic constraint from input causing a crash.
These constraints could be from testing target functions or during backtracking.
We also perform constraint reduction, an optimization to remove unnecessary variables.

\vspace{0.3em}
\indent\textbf{Method.}
We evaluate the robustness of this component by measuring the crashes for which constraint generation was successful.
We validate the accuracy of these constraints by random sampling.
We also measure the reduction rate of our constraint reduction technique.

\vspace{0.3em}
\indent\textbf{Results.}
\ourtool{} generated a total of \rqoneTotalCrashes{} unique target function crashes, and successfully generated \rqoneTotalConstraintsGenerated{} constraints (\rqoneTotalConstraintsGeneratedPercentage{}) for these crashes.
For call edges, we found \rqoneCallEdgeCrashes{} unique crashing inputs and generated \rqoneCallEdgeConstraints{} (\rqoneCallEdgeCrashesPercentage{}).
In total we were able to reduce the size of the constraints by \rqoneTotalSizeReductionsPercentage{}.
We explain the reasons for the failures in ~\apdx{appendix:constraintfailures}.

\subsubsection{Constraint Stitching}
As mentioned in \sect{sec:stitcher}, this component stitches constraints of the callee with the caller to filter out false positives.

\vspace{0.3em}
\indent\textbf{Method.}
We evaluate the effectiveness by the number of successfully stitched constraints.
We also measure the accuracy of the stitched constraints.

\vspace{0.3em}
\indent\textbf{Results.}
Our technique \emph{was able to successfully stitch all the constraints and provide either satisfiable or unsatisfiable stitched constraints for each case}.
To measure the accuracy of the stitching constraints, we picked all true positive cases and verified that the constraint contained both the root cause of the crash, and the path constraints for each of the target functions in the call chain.
We also randomly sampled 100 unsatisfiable stitching cases; for each case, we extracted the unsat-core using \emph{z3}.
We found that there were no false negatives in the stitching process, and all unsat-cores \cite{marques2008algorithms} contained a constraint from the project, which was indeed unsatisfiable.

\begin{tcolorboxfloat}[htbp]
\noindent\fcolorbox{black}{green!30!white}{
  \parbox{\dimexpr\linewidth-3\fboxsep-2\fboxrule\relax}{\parskip=1pt\parindent=0pt
RQ2 results show that all components of \ourtool{} are independently effective in tackling the corresponding tasks. 
}
}
\end{tcolorboxfloat}

\subsection{RQ3: Discovering Unknown Vulnerabilities}

We further evaluate the capability of \ourtool{} to discover previously unknown vulnerabilities in real-world software.

\vspace{0.3em}
\indent\textbf{Dataset.}
For this experiment, we selected challenging hard-to-fuzz projects.
These are widely used, complex projects that lack existing fuzzing harnesses, often due to intricate codebases, absence of clear entry points, or non-standard input formats.
\tbl{tab:realworldataset} provides a breakdown of the dataset.

\begin{listing}
\begin{minted}{c}
// lib/libalpm/sandbox.c
bool _alpm_sandbox_process_cb_download(
            alpm_handle_t *handle,
            int callback_pipe) {
  ...
  size_t filename_size, cb_data_size;
  ...
  ASSERT(read_from_pipe(callback_pipe, 
    &filename_size, sizeof(filename_size)) != -1, 
    return false);

  MALLOC(filename, |\textcolor{yellow}{\faWarning}|filename_size + 1, return false);
    
  ASSERT(|\textcolor{red}{\faBug}|read_from_pipe(callback_pipe, 
    filename, filename_size) != -1,
    FREE(filename); return false);
    filename[filename_size] = '\0';
  ...
}
\end{minted}
  \caption{Vulnerability in the Pacman package manager.}
\label{lst:pacmanbug}
\end{listing}

\vspace{0.3em}
\indent\textbf{Setup.}
All experiments were conducted on the same infrastructure as in \sect{sec:rq1}.
We used extended budgets for both fuzzing and symbolic execution runs, capped at 24 hours.

\vspace{0.3em}
\indent\textbf{Method.}
We prioritized functions based on the target identification and also prioritized functions that appeared security-critical or exposed to untrusted input.
If backtracking failed at any point, we stitched constraints up to the last feasible edge and manually analyzed the resulting bug reports. 
This approach ensures that limitations in the fuzzer or symbolic engine do not mask true vulnerabilities.

\vspace{0.3em}
\indent\textbf{Results.}
\ourtool{} was able to successfully identify 6 previously unknown vulnerabilities in these projects (\tbl{tab:realworldataset}).
We reported all the vulnerabilities to the respective maintainers.
Four of these vulnerabilities have already been patched, while the others are still under review.

\begin{table}[]
\caption{Summary of the dataset for finding new vulnerabilities}
\resizebox{\columnwidth}{!}{
\begin{tabular}{c|c|c|c}
\hline
\textbf{Target Name} & \textbf{Type}   & \textbf{SLOC} & \textbf{\begin{tabular}[c]{@{}c@{}}New\\ Vulnerabilities\end{tabular}} \\ \hline
mblaze               & Email Client    &      8,744         &    1                                                                    \\ \hline
pacman               & Package Manager &       23,898        &     4                                                                   \\ \hline
pspg                 & Unix Pager      &     23,767          &     1                                                                   \\ \hline
\end{tabular}
}
\label{tab:realworldataset}
\end{table}

\noindent\lst{lst:pacmanbug} shows a exploitable real-world vulnerability that we found and fixed in the Pacman package manager \cite{pacman}. 
This occurs due to a missing check on the \code{filename_size} variable. If the value provided for the variable is $2 ^{64} - 1$, the maximum possible value for that variable, then the subsequent addition at line marked with \textcolor{yellow}{\faWarning} will cause it to overflow and become $0$.
 Since the default behavior of \code{malloc} is to return a valid memory when the size is 0, the read at line marked with \textcolor{red}{\faBug} will try to read in $2 ^{64} - 1$, causing a heap overflow.
 \apdx{apnx:realbugs} contains an additional example.

\begin{tcolorboxfloat}[htbp]
\noindent\fcolorbox{black}{green!30!white}{
  \parbox{\dimexpr\linewidth-3\fboxsep-2\fboxrule\relax}{\parskip=1pt\parindent=0pt
RQ3 results show that \ourtool{} is capable of finding critical, previously unknown vulnerabilities in popular applications.
}
}
\end{tcolorboxfloat}

\subsection{Limitations and Future Work}
\label{sec:limitations}

This section discusses the limitations of \ourtool{} and potential future work to address these limitations.

\vspace{0.3em}
\indent\textbf{Concrete Input Generation:}
Currently, our implementation is unable to reliably generate concrete inputs(concretization) at the program entry point that exercise the full path to a target function.
This difficulty arises primarily from two factors: (1) Programs often accept input from multiple sources(files, sockets, and arguments etc.) requiring the tool to synthesize and coordinate data across these channels, which is non-trivial.
(2) For inputs involving arrays passed between multiple functions, there might be missing constraints that hinder concretization but don't hinder the stitching process.

\vspace{0.3em}
\indent\textbf{Fuzzing Performance:}
Our current fuzzing approach struggles to achieve high coverage when targeting top-level functions (e.g., \texttt{main}), particularly in programs that expect highly structured or complex inputs (such as specific file formats).
Our modular design allows for the integration of advanced fuzzing strategies, such as grammar-based fuzzing or file format fuzzers, which could significantly improve coverage.

\vspace{0.3em}
\indent\textbf{Support for Complex Recursive Data Structures:}
The type inference and harness generation components may not fully capture the semantics of complex or interdependent data structures.
We consider each pointer independently and fail to generate cyclic data structures, \eg{} cyclic linked lists.
Future work could incorporate more sophisticated type and relationship inference to better model such structures.
\vspace{-0.5em}

\section{Related Work}
\label{sec:related_work}

\hspace*{0.9em} 
\textbf{Hybrid Testing: }
There have been several attempts to combine different testing techniques into hybrid techniques~\cite{arbiter, parvez2016combining,shastry2017static} to improve the overall effectiveness of bug detection.
Hybrid Concolic Testing~\cite{majumdar2007hybrid, stephens2016driller, qsym, sage} proposes combining fuzzing and symbolic execution, in which an application is first subjected to fuzzing until the fuzzer is unable to find new paths, after which symbolic execution is used to trigger new paths.
However, unlike \bout{}, these approaches operate at the application level, and do not focus on specific components of interest.
Also, they require substantial computational resources when applied to evolving codebases, as even minor changes necessitate re-analyzing large portions of the application to ensure correctness.

\vspace{0.3em}
\indent\textbf{Directed Testing: }
Directed testing approaches can also be considered related to \bout{}, since they aim to guide the testing process towards specific components of interest.
Existing Directed Greybox fuzzing approaches can be classified into, (1) seed and mutation prioritization~\cite{bohme2017directed, 10.1145/3238147.3238176}, and (2) static analysis-based path pruning~\cite{fuzzingPrashast, 9833751}.
However, if the target is a function deep within the program, significant resources may be spent generating inputs that merely reach it.
Similarly, if multiple targets are specified, the exploration may be redundant.    
Directed symbolic execution has also been proposed to address similar challenges~\cite{ma2011directed, yang2014directed}, but these approaches suffer from the same limitations, and are often limited to research prototypes.
In summary, existing Directed Greybox testing techniques face challenges in efficiently and systematically targeting arbitrary internal functions or multiple points of interest within large codebases.
In contrast, \bout{} is designed to directly test specific functions of interest, enabling precise analysis and testing without the need for extensive static analysis or path pruning.

\section{Conclusions}

We present Reactive \bout{}, a novel testing paradigm that enables targeted testing from arbitrary program points, such as specific functions of interest. We realize this approach with \ourtool{},
which combines fuzzing to discover crashes, symbolic execution to generate constraints for them, and performs call graph backtracking to assess whether discovered vulnerabilities are reachable from higher-level entry points.
Using \ourtool{}, we discovered six previously unknown vulnerabilities across three real-world applications, demonstrating the effectiveness of bottom-up testing in uncovering software flaws.
\cleardoublepage
\appendix

\section{Acknowledgments}
This work was supported in part by Rolls-Royce and by the US National Science Foundation (NSF) under Grant CNS-2340548.
Any opinions, findings, conclusions, or recommendations expressed in this material are those of the author(s) and do not necessarily reflect the views of the Rolls Royce and NSF.

\section*{Ethical Considerations}
We followed the guidance of Davis \etal \cite{davis2025guidestakeholderanalysiscybersecurity} and evaluated the ethical considerations of our research by examining the potential impact on both direct and indirect stakeholders.
The most immediate stakeholders are software developers and maintainers, who may need to address vulnerabilities uncovered through our framework.
To address this, we followed responsible disclosure practices. The research team itself also bears responsibility: while developing and publishing this framework contributes to the advancement of software testing and security research, we recognize the professional risks associated with vulnerability discovery, and we have taken steps to ensure the ethical handling and communication of our results.

Our \ourtool{} framework can be used to detect defects, either with the intent of protecting users or exploiting systems. This makes adversaries another direct stakeholder. We do not provide proof-of-concepts for the discovered vulnerabilities to minimize the risk of their being used in an adversarial manner.
Indirect stakeholders include end users and society at large, who ultimately rely on the security of these systems.
Our approach aims to strengthen, rather than weaken, this security by enabling earlier detection of flaws.
We believe that advancing automated methods for vulnerability discovery, when handled responsibly, contributes positively to the public interest.

\bibliographystyle{plain}
\bibliography{griller}

\appendix

\section{Appendix}
\label{appendix:background}

\subsection{Outline of Appendices}

\noindent
The appendix contains the following material:

\begin{itemize}[leftmargin=12pt, rightmargin=5pt]

\item \cref{apdx:implementation}: Summary of Implementation
\item \cref{apdx:metrics}: Function Identification Metrics
\item \cref{appendix:opaque_types}: Handling Special cases and Opaque types 
\item \cref{apdx:runtimehooks}: Runtime Hooks
\item \cref{apdx:typeprinter} Type Printer
\item \cref{apdx:reduction} Symbolic Constraintss 
\item \cref{appendix:dataset-filtering} Dataset filtering
\item \cref{apnx:realbugs} Real World Bugs found by \ourtool{}
\item \cref{appendix:failureanalysis} Failure Analysis
\end{itemize}

\subsection{Implementation}
\label{apdx:implementation}
We implemented \ourtool{} as a modular framework that integrates existing state-of-the-art tools with our custom components to enable \bout testing.
We used LLVM~\cite{LLVM} version 10.0.0 as the underlying compiler infrastructure to perform code instrumentation and analysis.
We extended {\sc 3c}~\cite{3c} to implement our type inference and correlation analysis.
Our function fuzzer is implemented by extending AFL++~\cite{AFLplusplus-Woot20}.
All symbolic execution and constraint components are implemented using KLEE~\cite{klee} and Z3~\cite{z3}.
In total, our implementation includes 29,743 LoC of C/C++ and 28,432 SLoC of Python code.

\subsection{Metrics}
\label{apdx:metrics}

\subsubsection{Vulnerability Metrics}
We draw inspiration from the vulnerability metrics proposed by Leapord~\cite{leapord}.
However, since Leapord is a closed source tool, we implemented all the metrics proposed by Leapord.
The metrics we implemented are:
\begin{enumerate}
  \item Number of parameter variables
  \item Number of variables as parameters for callee function
  \item Number of pointer arithmetic
  \item Number of variables involved in pointer arithmetic
  \item Maximum pointer arithmetic a variable is involved In
  \item Number of Nested Control Structures
  \item Maximum nesting level of control structures
  \item Maximum of control-dependent control structures
  \item Maximum of data-dependent control structures
  \item Number of if structures without else
  \item Number of variables involved in control predicates
\end{enumerate}
We calculate each score and then normalize each individual metric across the entire target.
We then calculate the sum of all the normalized scores.

\subsubsection{Fuzzability Metrics}
\label{apdx:fuzzmetrics}
\algo{algo:fuzzability} shows the pseudocode of our Type Scoring algorithm.
The only special cases during the generation of the type score are the presence of pointers to the same type in structs, usually indicating that it's a linked list or some complex data structure. 
However, while \ourtool{} can generate these types, it's not aware of the specific data structure and the relationships between elements within it.
Note that while some input spaces will be marked as hard to generate,~\ourtool{} can still generate proper inputs for them, just that they might be harder to find paths for and to mutate.
In these cases, we can either go up or down the call graph to find more suitable functions for function-level testing.

\subsubsection{Ranking}
Finally, we incorporate the fuzzability metric to further filter the functions using the following formula:
\[
\text{score}_{\text{func}} = \text{vulnerability} \times \left( \frac{\text{fuzzability}_{\text{largest}}}{\text{fuzzability}_{\text{func}}} \right)
\]
where $\text{fuzzability}_{\text{largest}}$ is the largest fuzzability value from the list of target functions.
We then perform a binning technique similar to Leapord to group functions into bins based on size.
Instead of having a fixed number of bins, we dynamically increase the number of bins based on the number of functions that went into the last bin.
This is because larger functions tend to accumulate in the last bin if our initial bin size was too small.
Therefore, adding new bins when the functions in the last bin reach a threshold will make sure that no bin is overpopulated.
We then pick the top \emph{k} functions from each bin.
The value of \emph{k} depends on the number of functions in the bin.
The selected functions are marked as HIGH PRIORITY and are ordered based on their score.

We also select the bottom set of functions from each bin that have scores less than our threshold.
This threshold is selected such that one-third of the functions get classified as LOW PRIORITY.
We set the low threshold to be \texttt{0.5}.
These functions are marked as LOW PRIORITY.
The remaining functions in each bin are marked as MEDIUM PRIORITY.

\begin{algorithm}[t!] 
    \footnotesize
    \caption{Algorithm for Fuzzability Score Calculation} 
    \label{algo:fuzzability}
    \begin{algorithmic}[1]
        \Require{Target type \texttt{targetType}}
        \Ensure{Complexity score for the given type}
        \vspace{1mm}
        \Function{TypeFuzzabilityScore}{\texttt{targetType}}
            \If {\Call{IsBaseType}{\texttt{targetType}}}
                \State \Return 1
            \ElsIf {\Call{IsStructType}{\texttt{targetType}}}
                \State score $\gets$ 0
                \If {\Call{HasSelfPointer}{\texttt{targetType}}}
                    \State \Return $\infty$ \Comment{Mark as hard to generate}
                \EndIf
                \ForAll {\texttt{element} $\in$ \Call{ElementsOf}{\texttt{targetType}}
                    \State score $\gets$ score + \Call{TypeFuzzabilityScore}{\texttt{element}}}
                \EndFor
                \State \Return score
            \ElsIf {\Call{IsArrayType}{\texttt{targetType}}}
                \State \texttt{arraySize} $\gets$ \Call{DetermineSize}{\texttt{targetType}}
                \If {\texttt{arraySize} = unknown}
                    \State \texttt{arraySize} $\gets$ 128 \Comment{Average size for unknown arrays (range 0 - 255)}
                \EndIf 
                \State \texttt{arrayElement} $\gets$ \Call{ElementType}{\texttt{targetType}}
                \State \texttt{elementScore} $\gets$ \Call{TypeFuzzabilityScore}{arrayElement}
                \State \Return \texttt{elementScore} $\times$ \texttt{arraySize}
            \ElsIf {\Call{IsPointerType}{\texttt{targetType}}}
                \State \texttt{pointerElement} $\gets$ \Call{ElementType}{\texttt{targetType}}
                \State \Return \Call{ScoreTypeComplexity}{\texttt{pointerElement}} $\times$ 2
            \EndIf
        \EndFunction
    \end{algorithmic}
\end{algorithm}

\begin{algorithm}[t!] 
    \footnotesize
    \caption{Harness Generation Algorithm} 
    \label{algo:typegen}
    \begin{algorithmic}[1]
        \Require{Target type \texttt{targetType}}
        \Ensure{Generated code for the given type}
        \vspace{1mm}
        \Function{GenerateType}{\texttt{targetType}}
            \State typeSize $\gets$ \Call{SizeOf}{\texttt{targetType}}
            \If {\Call{IsBaseType}{\texttt{targetType}}}
                \State \Call{InsertBaseTypeGenerator}{\texttt{targetType}}
            \ElsIf {\Call{IsStructType}{\texttt{targetType}}}
                \State \texttt{struct} $\gets$ \Call{AllocateMemory}{\texttt{typeSize}}
                \ForAll {\texttt{element} $\in$ \Call{ElementsOf}{\texttt{targetType}}}
                    \State \Call{GenerateType}{\texttt{element}}
                \EndFor
            \ElsIf {\Call{IsArrayType}{\texttt{targetType}}}
                \State \texttt{arraySize} $\gets$ \Call{DetermineSize}{\texttt{targetType}}
                \State \texttt{arrayElement} $\gets$ \Call{ElementType}{\texttt{targetType}}
                \State \Call{GenerateLoopHeader}{\texttt{arraySize}}
                \State \Call{GenerateType}{\texttt{arrayElement}}
                \State \Call{GenerateLoopTerminator}{\texttt{arraySize}}
            \ElsIf {\Call{IsPointerType}{\texttt{targetType}}}
                \State \Call{GenerateNullOption}{}
                \State \texttt{pointer} $\gets$ \Call{AllocateMemory}{\texttt{typeSize}}
                \State \texttt{pointerElement} $\gets$ \Call{ElementType}{\texttt{targetType}}
                \State \Call{GenerateType}{\texttt{pointerElement}}
            \EndIf
        \EndFunction
    \end{algorithmic}
\end{algorithm}

\subsection{Special cases and Opaque Types}
\label{appendix:opaque_types}
We special case the following types:
\begin{itemize}
    \item \textit{Pointers:} For pointers, we must also handle the possibility of null pointers. 
    To address this, \ourtool{} generates a dedicated \emph{null option byte} for each pointer-type argument, which we notate to as \code{<arg>_nullopt}.
    This byte is read from the input and determines whether the pointer is set to \code{NULL} or a valid address, we set the pointer to \code{NULL} if the byte is less than 13(5\% probability), and to a valid address otherwise.
    This is illustrated in \code{gen_null} function in \lst{lst:trampoline}.
    \item \textit{Arrays:} If the size of the array is determined during the Type Identification phase, that size is used for the loop. 
    If the size is unknown, we ask the fuzzer to generate an unsigned integer to be used as the size.
    \item \textit{Function Pointers:} 
    Call signature matching similar to what we did for \sect{designone:indirectcall} is used to identify potential targets of the function pointer.
    A list of possible values is generated, and a fuzzer-generated integer determines which function pointer is used.
\end{itemize}

Some projects often utilize \emph{opaque types}, which are types not defined within the application code.
Such types typically originate from external libraries integrated into an application.
For example, in the tcpreplay~\cite{tcpreplay} project, the \texttt{struct  pcap\_t} type is an opaque type defined in the \texttt{libpcap} library.
In the LLVM IR for the tcpreplay project, this type appears as: \code{struct.pcap_t = type opaque}.

While the fields of this type are not directly accessed by the application code, such opaque types can be frequently used in the application to be passed as arguments to functions from the corresponding library.
The absence of a concrete type definition presents a significant challenge for harness generation, as the initialization procedure remains undefined.
To address this challenge, \ourtool{} implements a mechanism allowing users to specify custom type-harnesses written in C for these opaque types.
\lst{lst:opaque_harness} demonstrates a harness for the \texttt{pcap\_t} type in the tcpreplay project.
In this implementation, the struct is initialized using the \code{pcap_open_offline} function, which requires a filename as its argument.
Since \code{pcap_open_offline} returns a pointer to the struct, only in the presence of a valid pcap file, we need to provide a valid pcap file as the seed.
An important point to note is that the functions utilizing this struct pointer within the target function will continue to read data from stdin rather than from the initialization file, due to the hooks implemented by \ourtool{}.
The seed file serves solely to initialize the struct and can be any file satisfying the validation requirements of the \code{pcap_open_offline} function.

\begin{listing}
\begin{minted}{c}
pcap_t* CreatePcapTType(char *VN, int nargs, ...) {
    char errbuf[PCAP_ERRBUF_SIZE];
    pcap_t* handle = pcap_open_offline("/seeds/test.pcap",
                                       errbuf);
    if (!handle) {
        fprintf(stderr, "Failed to open PCAP file: %s\n",
                errbuf);
        return NULL;
    }
    return handle;
}
\end{minted}
  \caption{Harness written for an Opaque Type in tcpreplay - \emph{struct pcap\_t}.}
\label{lst:opaque_harness}
\end{listing}

\subsection{Runtime Hooks}
\label{apdx:runtimehooks}

These hooks redirect all external input operations to instead read from stdin, where the fuzzer has complete control over the data stream.
For example, when fuzzing the \code{handle_req} function in \lst{lst:runexample} (c), which reads data from \code{sockfd} into \code{rbuf} at Line 10, our runtime hook intercepts the \code{read} function call and substitutes the socket read operation with a stdin read operation.

The functions for which we generated runtime hooks are: \texttt{read}, \texttt{readv}, \texttt{fread}, \texttt{write}, \texttt{writev}, \texttt{fwrite}, \texttt{fprintf}, \texttt{fscanf}, \texttt{vfscanf}, \texttt{getenv}, \texttt{fopen}, \texttt{getc}, \texttt{fgetc}.
All hooks are implemented for both the KLEE and fuzzing runtimes to ensure execution parity.
Adding additional runtime hooks is very straightforward, and it can be achieved by adding the function definition to a file.
The runtime hooks will automatically be added to the target function.

\subsubsection{Klee Hooks}
Additionally, we instrument these functions with root cause assertions to capture relevant constraints on the symbolic input.
This approach ensures that all input-dependent behaviors observed during fuzzing are accurately reflected during symbolic execution, enabling precise constraint extraction. %

\subsection{Type Printer Details}
\label{apdx:typeprinter}
The \emph{TypePrinter} mirrors the structure of the \emph{TypeGenerator} used in the Function Harnesser: while the TypeGenerator constructs input values for fuzzing, the TypePrinter deconstructs arguments into their constituent fields and outputs their symbolic representations. 
This includes recursively handling complex types such as structs, arrays, and pointers, ensuring that all relevant subfields are captured.
An example of the TypePrinter instrumentation at the callsite of the \code{add_elem} function is shown in \lst{lst:type_printer} (in Appendix).
Note that we do not need to extract global variable values, as these are shared across the program and automatically included in the symbolic state.

Each symbolic value is assigned a unique and consistent name, following a naming convention that encodes the argument position and field position (in case of structures).
This naming scheme is critical for later constraint stitching, as it allows the Stitcher to match arguments between parent and child functions unambiguously.
To extract the values during execution inside KLEE, we extend KLEE with custom API functions (such as \code{grill_print_<type>}) to emit the symbolic values in SMT-LIB~\cite{smt-lib} format.

When dumping argument values at the callsite (for constraint stitching), we encode the nullness using the same convention:
\begin{itemize}
    \item If the pointer is \code{NULL}, we dump \code{255} for its null option byte.
    \item If the pointer is not \code{NULL}, we dump \code{0}.
\end{itemize}
This allows us to match the nullness of pointers between caller and callee during constraint stitching.

\begin{listing}
\begin{minted}[breaklines,fontsize=\scriptsize,escapeinside=||]{c}
int process_req(enum rtype r, 
                uint32_t idx,
		char *buf, 
		uint32_t sz) {

 if (idx < POOL_SIZE) {
   GP[idx].size = size;
   switch(r) {
     case UPDATE:
|\text{\faArrowRight}|      print_ptr_to_pool(&GP[idx]);
|\text{\faArrowRight}|      print_char_ptr_with_size(buf, sz);
|\text{\faArrowRight}|      debug_trap();
       add_elem(&GP[idx], buf, sz);
       break;
     ...    
   }
 }
 ...
}

void print_ptr_to_pool(struct pool *p) {
    if (check_null("p", p)) return;
    grill_print("p->size", p->size);   // print
    ....
}

void print_char_ptr_with_size(char *buf, uint32_t sz) {
    if (check_null("buf", buf)) return;
    grill_print_array_size("sz", sz);
    for (int i = 0; i < sz; i++)
        grill_print_char("buf[i]", buf[i]); // print
    ....
}
\end{minted}
\caption{TypePrinters inserted at the callsite of \code{add_elem} in \code{process_req}.~\faArrowRight~indicates printers that were added.}
\label{lst:type_printer}
\end{listing}

\subsubsection{Runtime Hooks:}
As discussed in \emph{Challenge 2}, it is essential that the runtime hooks used during fuzzing are faithfully reproduced during symbolic execution to maintain consistency in input handling.
To achieve this, we allocate a symbolic buffer, \code{stdin_buf}, of configurable size (\code{stdin_size}), which serves as the source for all standard input operations within KLEE. 
Similar to \sect{sec:funcfuzz}, standard library functions are modified to use \code{stdin_buf} buffer rather than the original source.
Note that dynamic hooking of these functions (\eg{}~\code{LD_PRELOAD}) does not work as KLEE works directly on the LLVM IR.

\subsection{Symbolic Constraints}
\label{apdx:reduction}

\subsubsection{Symbolic Sizes and Null Pointers}
\label{apdx:symbolic_sizes}
\paragraph{Symbolic Sizes}
To address this, we implement a runtime tracking mechanism that records the symbolic size associated with each dynamically allocated array.
Specifically, before every memory allocation (e.g., \code{malloc}, \code{calloc}), we ensure that the size argument is made concrete.
After allocation, we store a mapping from the returned pointer to its symbolic size. 
This mapping is maintained throughout execution and is referenced whenever assertions are added at memory access sites (e.g., buffer overflows, out-of-bounds accesses) by the \emph{Assertion Adder}.
\lst{lst:kleesymbolicsize} illustrates our KLEE runtime hooks for tracking symbolic sizes.

\begin{listing}
\begin{minted}{c}
int add_elem(void *data, 
             const char *buf, 
             unsigned size) {
 struct pool *p = (struct pool *)data;
 if (p->avail) {
    size_t temp = concretize_size(p->size *
                                  sizeof(char));
    p->buf = malloc(temp);
    klee_track_pointer(p->buf, p->size * sizeof(char));
    if (p->buf) {
      p->avail = 0;
      memcpy(p->buf, buf, size);
    }
   return 0;
 }
 return -1;
}
\end{minted}
  \caption{KLEE Runtime hooks for symbolic sizes.}
\label{lst:kleesymbolicsize}
\end{listing}

\paragraph{Null Pointer Constraints}
As discussed in \emph{Challenge 3}, \ourtool{} needs to generate constraints to represent that a pointer is null.
\ourtool{} doesn't generate the arrays in a symbolic manner, so we can't generate the constraints for the null pointer.
To workaround this, we take advantage of the \emph{null option byte} (\sect{sec:harnessgen}) that we generate for each pointer-type argument.
By making this byte symbolic, we can generate a constraint to determine whether a pointer is \code{NULL} or not.

Consider the \code{add_elem} function from \lst{lst:runexample} (a), which takes two pointers: \code{data} and \code{buf}. 
If the harness generates a null pointer for \code{data}, it is because the byte read from stdin, \code{data_nullopt} was less than 13.
This constraint \code{data_nullopt < 13} is used to determine whether the pointer \code{data} is \code{NULL} or not.
We will refer to this as the \emph{null option} constraint.

\subsubsection{Size Reduction}
We identified size variables based on the \texttt{input grammar}.
We used PySMT to parse constraints relevant to the size variables and then replaced them with a boolean \texttt{true} statement.
Since the constraints were in Conjunctive Normal Form (CNF), this did not affect the other constraints.

\subsubsection{Null Pointer Reduction}
To remove the null pointer constraints for all the null pointers that are generated, we perform the following steps.

\noindent \textbf{Target Functions}:
When analyzing Target Functions, it is important to determine whether null pointer constraints generated in the harness are actually relevant to a crash.
To achieve this, we systematically mutate the input that triggered the crash and observe the effect on program behavior.
After generating the path constraints for a crashing input, we identify all pointer variables introduced by the harness. 
For each such pointer, we create two mutated inputs: one where the pointer is forced to be null, and another where it is forced to be non-null (i.e., assigned a valid value). 
Formally, for each pointer $j$ in the input $I$, we generate a new input $I_j$ such that $pointer(j)$ is toggled between null and non-null.
Each mutated input $I_j$ is then executed using the driver program ($\dr$ ) to check whether the original crash is still reproduced.
If the crash occurs regardless of the pointer being null or non-null, we conclude that the null pointer constraint for that variable is not relevant to the crash. 
In such cases, we remove the corresponding constraint on the \emph{null option byte} for that pointer from the path constraints.

\noindent \textbf{Call Edges} :
For call edges, the process of removing irrelevant null pointer constraints is more nuanced due to the presence of two distinct constraint types: \emph{Path Constraints} (which capture the execution path within the $pf$) and \emph{Arg Constraints} (which encode the argument values at the callsite of $cf$).

For \emph{Path Constraints}, we apply the same mutation-based analysis as described above: for each pointer argument, we systematically toggle its nullness in the input and observe whether the crash is still reproduced. 
If the crash is unaffected by the pointer's value, we remove the corresponding null constraint from the path constraints.
However, for \emph{Arg Constraints}, the challenge lies in mapping the pointers in the arguments of $pf$ to the arguments passed to $cf$ at the callsite, as the relationship is not always direct.
To resolve this, we instrument the code to output a bitmap at each callsite, indicating which arguments are null during execution.
By correlating this bitmap with the mutated inputs, we can accurately determine which harness pointers correspond to which callsite arguments.
Once this mapping is established, any pointer found to be irrelevant for the path constraints (i.e., its nullness does not affect the outcome) has its associated constraint removed from the \emph{Arg Constraints} as well.
This ensures that only constraints essential to reproducing the crash are retained, improving the precision of our analysis.

\subsection{Dataset Filtering}
\label{appendix:dataset-filtering}

AFGen~\cite{AFGen} was evaluated on a dataset of 11 applications. However, we decided to perform our evaluation on a subset of the programs in the dataset. 
AFGen did not have a methodology for the selection of the dataset or for the selection of the CVEs inside the dataset. 

We have refined this dataset based on the following criteria:
\begin{enumerate}
    \item \textbf{Active maintenance:} 
    We excluded targets that have not had active commits in the last five years, ensuring our evaluation focuses on currently relevant software.
    For example, the \texttt{libwav}~\cite{libwav} project, was last updated in 2017, and all the CVEs reported were not verified, acknowledged or fixed by the developers.
    Similarly, the \texttt{libming}~\cite{libming} project was also last updated in 2020, and the project has over 200 open issues, most of which are related to memory corruption vulnerabilities.
    We decided to exclude projects that were not actively maintained, as they are less likely to be relevant for our evaluation.
    \item \textbf{Vulnerability verification:} We only included vulnerabilities that have been verified and fixed, enhancing the credibility of our results.
    \item \textbf{Reproducibility:} We ensured that all included vulnerabilities are reproducible, facilitating future research and validation of our findings.
\end{enumerate}

\paragraph{Side Effect Bug}
\label{apdx:side_effect}
During our analysis of ngiflib~\cite{ngiflib}, we observed a discrepancy between the execution paths taken during concrete (fuzzing) and symbolic execution.
Upon closer inspection, we identified that the function \code{GetByteStr} was invoked in \code{DecodeGifImg}, but its return value was not checked by the caller. 
This omission led to an uninitialized memory read at the location marked with \textcolor{red}{\faBomb} in \lst{lst:side_effect}.
Subsequent use of this uninitialized memory resulted in undefined behavior, causing the program to follow divergent paths.
This divergence complicated our analysis and could potentially mask real bugs or produce false positives.
We reported this issue to the ngiflib~\cite{ngiflib} developers, highlighting the importance of always validating the results of input-reading functions to ensure consistent and correct program behavior.
We patched these bugs that were not relevant to the target bugs.

\begin{listing}
\begin{minted}{c}
static int GetByteStr(struct ngiflib_gif * g,
                      u8 * p,
                      int n) {
    ...
	read = fread(p, 1, n, g->input.file);
	return ((int)read == n) ? 0 : -1;
}

static u16 GetGifWord(...,
                      struct ngiflib_context * context) {
    ...
	GetByteStr(i->parent,
                   context->byte_buffer,
                   context->restbyte);|\textcolor{red}{\faBug}|
	context->srcbyte = context->byte_buffer;
    ...
    r = *context->srcbyte++;|\textcolor{red}{\faBomb}|
}

static int DecodeGifImg(struct ngiflib_img * i) {
    ...
|\text{\faArrowRight}|    struct ngiflib_context context; 
    ...
    act_code = GetGifWord(i, &context);
}
\end{minted}
  \caption{Bug identified in ngiflib~\cite{ngiflib} during KLEE failure analysis.}
\label{lst:side_effect}
\end{listing}

\subsection{Real World Bugs}
\label{apnx:realbugs}
\begin{listing}
\begin{minted}{c}
// mblaze/seq.c

long indent = 0;
while (*s && iswsp(*s)) {
    s++;
    indent++;
}
if (indent > 255)
    indent = 255;
previndent[indent] = line;
if (line == *starto) {
    if (|\textcolor{red}{\faBug}|previndent[indent-1]) {
    ...
    }
}
\end{minted}
  \caption{Out of Bounds array access in the mblaze.}
\label{lst:mblazebug}
\end{listing}

\lst{lst:mblazebug} shows an OOB access in mblaze \cite{mblaze} that is caused due to a missing check on the \code{indent} variable.
If \code{s} does not have any white spaces, the array access at the line marked with \textcolor{red}{\faBug} will access index $-1$, possibly resulting in undefined behavior.

\subsection{Failure Analysis}
\label{appendix:failureanalysis}
In this section, we dive deeper into the reasons \ourtool{} was unable to find some of the bugs or failed to backtrack it properly.
We also cover individual component level failures to validate the results in the papers, however most individual failures might not have affected vulnerability discovery as a whole

\subsubsection{False Negatives in Backtracking}
Backtracking may produce false negatives when attempting to generate valid inputs that traverse from high-level functions to specific call sites, particularly in functions with strict input validation requirements.

Consider the snippet~\lst{lst:jiff_challenge}, where we are trying to reach \code{jfif_decode} from \code{main}. 
The intermediate function \code{jfif_load} performs input validation and returns \code{NULL} for malformed files.
Since generating syntactically valid image files is hard, the fuzzer predominantly produces malformed inputs. 
Consequently, the generated inputs result in \code{NULL} arguments being passed to \code{jfif_decode}.

However, \code{jfif_decode} contains defensive programming checks that explicitly handle \code{NULL} inputs, causing early termination.
This creates a mismatch and the resulting symbolic constraints become unsatisfiable, preventing us from validating the vulnerablities found inside \code{jfif_decode}.

\begin{listing}
\begin{minted}{c}
int jfif_decode(void *ctxt, BMP *pb)
{
    ... 
    if (!ctxt || !pb) {
        printf("invalid input params !\n");
        return -1;
    }
}

void* jfif_load(char *file)
{
    JFIF *jfif   = NULL;
    // tries to load a valid jfif file
    ...
    
    // if file has some invalid field
    if (ret == -1) {
        jfif_free(jfif);
        jfif = NULL;
    }
    return jfif;
}

int main(int argc, char *argv[])
{
    ...
    jfif = jfif_load(argv[2]); 
    jfif_decode(jfif, &bmp);
}
\end{minted}
  \caption{Example showing input generation challenge.}
\label{lst:jiff_challenge}
\end{listing}

\subsubsection{Reachability issues}
In about 10\% of the call-edges \ourtool{} was unable to generate a valid input that reached the callsite for the child function from the parent.
We mainly noticed 3 patterns for this happening -
\begin{enumerate}
    \item \textbf{Complex input format:} similar to the example discussed in the previous section, the fuzzer requires a specific input format.
    \item \textbf{Unmodeled Environment:} functions from third-party libraries such as libpcap, which check for availability of network devices prohibit fuzzer from proceeding further.
    \item \textbf{Crashes:} In certain cases, there are crashes in the parent function that prevent it from reaching the child function. This is mainly due to the fact that our strict runtime santization does't allow the fuzzer to proceed from these crashes. In some of the cases we patched these unrelated vulnerabilities. 
\end{enumerate}

\subsubsection{Reachability Limitations}
\ourtool{} encountered reachability issues in approximately 10\% of call-edges, where valid inputs could not be generated to traverse from parent to child functions.
We identified three primary patterns causing these limitations:
\begin{enumerate}
    \item \textbf{Complex Input Formats:} As demonstrated in the previous section, functions requiring strictly structured input formats (e.g., valid image files, network protocols) present significant challenges for constraint-based input generation.
    \item \textbf{Unmodeled Environmental Dependencies:} In certain cases, functions utilizing third-party libraries (e.g., libpcap) often include runtime checks for system resources such as network device availability or file system permissions. 
    If these are unmodeled, execution terminates before reaching target functions.
    \item \textbf{Intermediate Crashes:} 
    Parent functions containing unrelated vulnerabilities prevent execution from reaching child functions due to our strict runtime sanitization policies.
\end{enumerate}

\subsubsection{Constraint Generation Failures}
\label{appendix:constraintfailures}

There were no Constriant Generation failures that affected all the inputs in an call-edge or a target functions.
Failures were limited to certain set of inputs in that group.
Our investigation into constraint generation failures identified two primary causes. 
\begin{enumerate}
    \item \textbf{KLEE memory errors:}
    KLEE may encounter memory errors (due to invalid access) during execution, this is due to limitation of ASan, which cannot track off-by-one errors and cannot validate pointers that are coming from external libraries.
    However, in most cases, \ourtool{} (while fuzzing) is able to generate another valid input that does not cause a memory error.
    \item \textbf{Path Disparity:}
    While GrillSan is able to handle most of the cases where the path diverges, there are cases where the path diverges and KLEE is unable to reach the callsite of the callee function.
    The two most common reasons that we observed for this are:
    (1) Floating point operations: KLEE does not support floating point operations, and tries to patch the seed input causing the path to diverge.
    (2) Pointer arithmetic: The memory model of KLEE is different, so any pointer operations such as difference between two pointers, and consequent usage of the result for control flow will cause KLEE to diverge.
    These are known issues with KLEE, and are not specific to our implementation.
\end{enumerate}

\begin{listing}
\begin{minted}{c}
int main(int argc, char *argv[]) {
    if (argc < 3) { ... }

    if (strcmp(argv[1], "-d") == 0) {
        jfif = jfif_load(argv[2]);
        jfif_decode(jfif, &bmp); |\faArrowCircleLeft|
        ...
    } 
    ...
    return 0;
}

void* jfif_load(char *file) {
    // open and load the jfif file
    ...
    
    // if load failed
    if (ret == -1) {
        jfif_free(jfif);
        jfif = NULL;
    }
    return jfif;
}
\end{minted}
  \caption{Example from ffjpeg\cite{ffjpeg} where we were able to reach the callsite for \code{jfif_decode}(marked by \faArrowCircleLeft) but we were unable to parse a valid \code{jfif_load} in \code{jfif_decode}, making the only value we got as NULL.}
\label{lst:example_reached_but_failed}
\end{listing}

\begin{listing}
\begin{minted}{c}
u32 GifIndexToTrueColor(struct ngiflib_rgb * palette,
                        u8 v) {
	return palette[v].b ...;  |\faBomb|
}

static void WritePixels(struct ngiflib_img * i, ...) {
	...
		for(j = (int)tocopy; j > 0; j--) {
			GifIndexToTrueColor(i->palette,
                                    *pixels++); |\encircle{1}|
		}
}

static int DecodeGifImg(struct ngiflib_img * i) {
	...
	if (...) {
	    i->palette = ngiflib_malloc(...);
	} else {
		i->palette = i->parent->palette;
	}
	i->ncolors = 1 << i->localpalbits;
	
	for(;;) {
	    ...	
		WritePixels(i, ...); |\encircle{2}|
		...
	}
}

int LoadGif(struct ngiflib_gif * g) {
	if(...) {
		g->palette = ngiflib_malloc(...)
	} else {
		g->palette = NULL; 
	}
	switch() {
		...
		g->cur_img->parent = g;
		DecodeGifImg(g->cur_img); |\encircle{3}|
	}
}
\end{minted}
  \caption{True positive that \ourtool{} was able to find that AFGen was not able to find.}
\label{lst:ngiflib_tp}
\end{listing}

\begin{table}[]
\setlength{\tabcolsep}{13pt}
\caption{Performance of the Target Function Testing on the dataset}
\label{tab:rq2_leaf_effectiveness}
\resizebox{\columnwidth}{!}{
\begin{tabular}{c|cc|cc}
\toprule
\multirow{2}{*}{\textbf{Target}} & \multicolumn{2}{c|}{\textbf{Crashes}}        & \multicolumn{2}{c}{\textbf{Constraints}}    \\ \cline{2-5} 
                        & \multicolumn{1}{c|}{\textbf{\raisebox{-0.5ex}{Total}}} & \textbf{\raisebox{-0.5ex}{Unique}} & \multicolumn{1}{c|}{\textbf{\raisebox{-0.5ex}{Total}}} & \textbf{\raisebox{-0.5ex}{Unique}} \\ \midrule
             ngiflib           & \multicolumn{1}{c|}{414}      &    128    & \multicolumn{1}{c|}{656}      &   107     \\ \hline
             ffjpeg          & \multicolumn{1}{c|}{163}      &     42   & \multicolumn{1}{c|}{268}      &     32   \\ \hline
             tcpreplay           & \multicolumn{1}{c|}{335}      &  81      & \multicolumn{1}{c|}{408}      &   57     \\ \hline
             jasper           & \multicolumn{1}{c|}{345}      &   98     & \multicolumn{1}{c|}{332}      &     58   \\ \hline
             luPng           & \multicolumn{1}{c|}{62}      &    19    & \multicolumn{1}{c|}{106}      &    18    \\ \bottomrule
\end{tabular}
}
\end{table}

\begin{table}[]
\setlength{\tabcolsep}{15pt}
\caption{Summary of the number of unique call edges, the number of reachable call edges, and the total number of call edges for each target.}
\label{tab:rq2_call_edges}
\resizebox{\columnwidth}{!}{
\begin{tabular}{c|c|cc}
\toprule
\multirow{2}{*}{\textbf{Target}} & \multirow{2}{*}{\textbf{\begin{tabular}[c]{@{}c@{}}Unique \\ Call Edges\end{tabular}}} & \multicolumn{2}{c}{\textbf{Reachability}}  \\ \cline{3-4}
                                 &                                                                                        & \multicolumn{1}{c|}{\textbf{\begin{tabular}[c]{@{}c@{}}Some\\ Callsites\end{tabular}}} & \textbf{\begin{tabular}[c]{@{}c@{}}All\\ Callsites\end{tabular}}                                \\ \hline
ngiflib                          & 19                                                                                    & \multicolumn{1}{c|}{19}                                                               & 19                                                                                        \\ \hline
ffjpeg                           & 9                                                                                    & \multicolumn{1}{c|}{8}                                                               & 8                                                                                         \\ \hline
tcpreplay                        & 102                                                                                    & \multicolumn{1}{c|}{93}                                                               & 91                                                                                       \\ \hline
jasper                           & 54                                                                                    & \multicolumn{1}{c|}{42}                                                               & 40                                                                                        \\ \hline
lupng                            & 30                                                                                    & \multicolumn{1}{c|}{28}                                                               & 26                             \\ \bottomrule
\end{tabular}
}
\end{table}

\begin{table}[t!]
\caption{Summary of common types used in harness generation.}
\centering
\resizebox{\columnwidth}{!}{
\begin{tabular}{c|l|l}
\toprule
\textbf{Common Types} & \multicolumn{1}{c|}{\textbf{Total Count}} & \multicolumn{1}{c}{\textbf{\begin{tabular}[c]{@{}c@{}}Average Per\\  Function\end{tabular}}} \\ \hline
Basic Types           &  16785                                         &  55                                                                                         \\ \hline
Structures            &  3714                                         &  12                                                                                         \\ \hline
Arrays     &      2788                                     &   9                                                                                         \\ \hline
Promoted Arrays     &      6862                                     &   20                                                                                    \\ \hline
Pointers        &  413                                         &  1                                                                                         \\ \hline
Function Pointers     &  1743                                         &  5                              \\ \bottomrule
\end{tabular}
}
\label{tab:commonTypes}
\end{table}

\end{document}